%% file: acl2021.tex
\documentclass[11pt,a4paper]{article}
\usepackage[hyperref]{acl2021}
\usepackage{times}
\usepackage{latexsym}

\usepackage{lipsum}
\newcommand\blfootnote[1]{%
  \begingroup
  \renewcommand\thefootnote{}\footnote{#1}%
  \addtocounter{footnote}{-1}%
  \endgroup
}

\usepackage{url}
\usepackage{float}
\usepackage{graphicx}
\usepackage{multirow}
\usepackage{multicol}
\usepackage{microtype}
\usepackage{subcaption}
\usepackage{enumitem,kantlipsum}
\usepackage[normalem]{ulem}
\usepackage{tabularx}
\usepackage{tcolorbox}

\usepackage{xspace}
\usepackage{fontawesome}
\usepackage{amssymb, amsmath}
\usepackage{pbox}
\usepackage{booktabs} 
\usepackage{fancyvrb}
\usepackage{verbatimbox}
\usepackage{tablefootnote}
\usepackage{colortbl}

\usepackage{xcolor, colortbl}
\definecolor{Gray}{gray}{0.9}
\usepackage{tikz}
\usetikzlibrary{decorations.pathreplacing}

\newcommand{\rashkincorpus}{\texttt{TSHP-17}}
\newcommand{\proppycorpus}{\texttt{QProp} }

\newcommand{\sq}{\faCheckSquare}
\newcommand{\cq}{\faCheck}

\usepackage{lscape}	
\usepackage{booktabs}
\newcommand{\msrc}{\textbf{ \emph{Source(s):} }}
\newcommand{\imgsrc}[2]{\href{#1}{Image #2}}

\newcommand{\ccsnd}[1]{\href{https://creativecommons.org/licenses/by-sa/2.0/deed.en}{License #1}}

\newcommand{\cctrdunprt}[1]{\href{https://creativecommons.org/licenses/by/3.0/deed.en}{License #1}}
\newcommand{\ccfrth}[1]{\href{https://creativecommons.org/licenses/by-sa/4.0/deed.en}{License #1}}
\newcommand{\ccfrtint}[1]{\href{https://creativecommons.org/licenses/by/4.0/deed.en}{License #1}}
\newcommand{\unplash}[1]{\href{https://unsplash.com/license}{License #1}}
\newcommand{\pixelbay}[1]{\href{https://pixabay.com/service/license/}{License #1}}
\newcommand{\public}[1]{\href{https://creativecommons.org/publicdomain/zero/1.0/}{License #1}}

\aclfinalcopy 
\def\aclpaperid{218} 

\title{SemEval-2021 Task 6:\\
Detection of Persuasion Techniques in Texts and Images}

\author{
Dimitar Dimitrov,\textsuperscript{\rm 1}  
Bishr Bin Ali,\textsuperscript{\rm 2} 
Shaden Shaar,\textsuperscript{\rm 3} 
Firoj Alam,\textsuperscript{\rm 3} 
\\{\bf Fabrizio Silvestri,\textsuperscript{\rm 4}  
Hamed Firooz,\textsuperscript{\rm 5} 
Preslav Nakov, \textsuperscript{\rm 3} 
and Giovanni Da San Martino\textsuperscript{\rm 6} 
}\\
\textsuperscript{\rm 1} Sofia University ``St. Kliment Ohridski'', Bulgaria,
\textsuperscript{\rm 2} King's College London, UK,\\
\textsuperscript{\rm 3} Qatar Computing Research Institute, HBKU, Qatar\\
\textsuperscript{\rm 4} Sapienza University of Rome, Italy,
\textsuperscript{\rm 5} Facebook AI, USA, 
\textsuperscript{\rm 6} University of Padova, Italy\\
\texttt{mitko.bg.ss@gmail.com, bishrkc@gmail.com}\hspace{5mm}\\
\texttt{\{sshaar, fialam, pnakov\}@hbku.edu.qa, mhfirooz@fb.com}\hspace{5mm} \\
\texttt{fsilvestri@diag.uniroma1.it,
dasan@math.unipd.it}
  \\}

\date{}

\begin{document}
\maketitle
\begin{abstract}
We describe \emph{SemEval-2021 task 6 on Detection of Persuasion Techniques in Texts and Images}: the data, the annotation guidelines, the evaluation setup, the results, and the participating systems. The task focused on memes and had three subtasks: (\emph{i})~detecting the techniques in the text, (\emph{ii})~detecting the text spans where the techniques are used, and (\emph{iii})~detecting techniques in the entire meme, i.e.,~both in the text and in the image. It was a popular task, attracting 71 registrations, and 22 teams that eventually made an official submission on the test set. The evaluation results for the third subtask confirmed the importance of both modalities, the text and the image. Moreover, some teams reported benefits when not just combining the two modalities, e.g., by using early or late fusion, but rather modeling the interaction between them in a joint model.
\end{abstract}

\input{sections/introduction}
\input{sections/related_work}
\input{sections/propaganda_techniques}
\input{sections/dataset}

\input{sections/evaluation_framework}

\input{sections/participants_systems}

\input{sections/conclusion}

\section*{Acknowledgments}

This research  part of the Tanbih mega-project,\footnote{\url{http://tanbih.qcri.org/}}
which is developed at the Qatar Computing Research Institute, HBKU, and aims to limit the impact of ``fake news,'' propaganda, and media bias by making users aware of what they are reading.

\section*{Ethics and Broader Impact}
\paragraph{User Privacy}

Our dataset only includes memes and it contains no user information.

\paragraph{Biases}

Any biases in the dataset are unintentional, and we do not intend to do harm to any group or individual. Note that annotating propaganda techniques can be subjective, and thus it is inevitable that there would be biases in our gold-labeled data or in the label distribution. We address these concerns by collecting examples from a variety of users and groups, and also by following a well-defined schema, which has clear definitions and on which we achieved high inter-annotator agreement.

Moreover, we had a diverse annotation team, which included six members, both female and male, all fluent in English, with qualifications ranging from undergrad to MSc and PhD degrees, including experienced NLP researchers, and covering multiple nationalities. This helped to ensure the quality. No incentives were provided to the annotators.

\paragraph{Misuse Potential}

We ask researchers to be aware that our dataset can be maliciously used to unfairly moderate memes based on biases that may or may not be related to demographics and other information within the text. Intervention with human moderation would be required in order to ensure this does not occur.

\bibliographystyle{acl_natbib}
\bibliography{bib/acl2021,bib/custom,bib/semeval21papers}

\newpage
\clearpage
\section*{Appendix}
\label{sec:appendix}
\appendix
\input{sections/supplemental_material}

\end{document}

%% file: sections/introduction.tex
\section{Introduction}\blfootnote{WARNING: This paper contains meme examples and wording that might be offensive to some readers.}

\label{sec:introduction}

Internet and social media have amplified the impact of disinformation campaigns. Traditionally a monopoly of states and large organizations, now such campaigns have become within the reach of even small organisations and individuals~\cite{ijcai2020-672}.

Such propaganda campaigns are often carried out using posts spread on social media, with the aim to reach very large audience. While the rhetorical and the psychological devices that constitute the basic building blocks of persuasive messages have been thoroughly studied~\cite{Miller,Weston2000,Torok2015}, only few isolated efforts have been made to devise automatic systems to detect them~\cite{Habernal2018,Habernal2018b,EMNLP19DaSanMartino}.

\begin{figure}[t]
    \centering
    \includegraphics[scale=0.17]{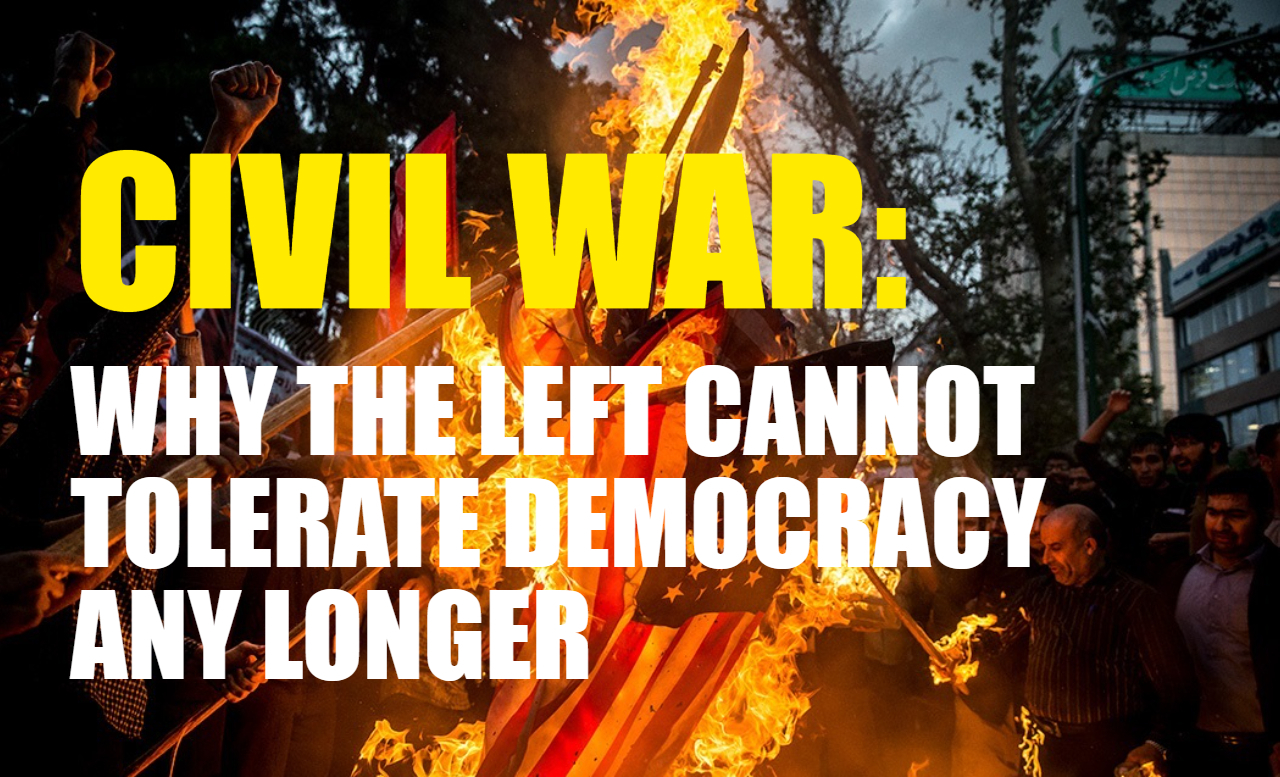}
    \caption{A meme with a civil war threat during the President Trump’s impeachment trial. Two persuasion techniques are used: (\emph{i})~\emph{Appeal to Fear} in the image, and (\emph{ii})~\emph{Exaggeration} in the text.
    \msrc
    \imgsrc{https://commons.wikimedia.org/wiki/File:Protests_after_US_decision_to_withdraw_from_JCPOA,_around_former_US_embassy,_Tehran_-_8_May_2018_25.jpg}{};
    \ccfrtint{}
    }
    \label{fig:meme_example}
\end{figure}

Thus, in 2020, we proposed \textit{SemEval-2020 task 11 on Detection of Persuasion Techniques in News Articles}, with the aim to help bridge this gap~\cite{da-san-martino-etal-2020-semeval}.
The task focused on text only. Yet, some of the most influential posts in social media use memes, as shown in Figure~\ref{fig:meme_example},\footnote{In order to avoid potential copyright issues, all memes we show in this paper are our own recreation of existing memes, using images with clear copyright.} where visual cues are being used, along with text, as a persuasive vehicle to spread disinformation~\cite{10.1145/3137597.3137600}. During the 2016 US Presidential campaign, malicious users in social media (bots, cyborgs, trolls) used such memes to provoke emotional responses~\cite{10.1145/3393880}. 

In 2021, we introduced a new SemEval shared task, for which we prepared a multimodal corpus of memes annotated with an extended set of techniques, compared to SemEval-2020 task 11. 
This time, we annotated both the text of the memes, highlighting the spans in which each technique has been used, as well as the techniques appearing in the visual content of the memes. 

Based on our annotations, we offered the following three subtasks: 

\begin{description}
\item \textbf{Subtask 1 (ST1)} Given the textual content of a meme, identify which techniques (out of 20 possible ones) are used in it. This is a multilabel classification problem.
\item \textbf{Subtask 2 (ST2)} Given the textual content of a meme, identify which techniques (out of 20 possible ones) are used in it together with the span(s) of text covered by each technique. This is a multilabel sequence tagging task.
\item \textbf{Subtask 3 (ST3)} Given a meme, identify which techniques (out of 22 possible ones) are used in the meme, considering both the text and the image. This is a multilabel classification problem.
\end{description}
 
A total of 71 teams registered for the task, 22 of them made an official submission on the test set and 15 of the participating teams submitted a system description paper.


%% file: sections/related_work.tex
\section{Related Work} 
\label{sec:related_work}

\paragraph{Propaganda Detection} Previous work on propaganda detection has focused on analyzing textual content \cite{BARRONCEDENO20191849,EMNLP19DaSanMartino,rashkin-EtAl:2017:EMNLP2017}. See \cite{da2020survey} for a recent survey on computational propaganda detection.
\citet{rashkin-EtAl:2017:EMNLP2017} developed the \rashkincorpus~corpus, which had document-level annotations with four classes: \emph{trusted}, \emph{satire}, \emph{hoax}, and \emph{propaganda}. Note that \rashkincorpus~was labeled using distant supervision, i.e., all articles from a given news outlet were assigned the label of that news outlet. The news articles were collected from the English Gigaword corpus (which covers reliable news sources), as well as from seven unreliable news sources, including two propagandistic ones. They trained a model using word $n$-grams, and reported that it performed well only on articles from sources that the system was trained on, and that the performance degraded quite substantially when evaluated on articles from unseen news sources.
\citet{BARRONCEDENO20191849} developed a corpus \proppycorpus~with two labels (propaganda vs. non-propaganda), and experimented with two corpora: \rashkincorpus~and \proppycorpus. They binarized the labels of \rashkincorpus~as follows: propaganda \textit{vs.} the other three categories. 

They performed massive experiments, investigated writing style and readability level, and trained models using logistic regression and SVMs. Their findings confirmed that using distant supervision, in conjunction with rich representations, might encourage the model to predict the source of the article, rather than to discriminate propaganda from non-propaganda. The study by \citet{Habernal.et.al.2017.EMNLP,Habernal2018b} also proposed a corpus with 1.3k arguments annotated with five fallacies, including \textit{ad hominem}, \textit{red herring}, and \textit{irrelevant authority}, which directly relate to propaganda techniques.

A more fine-grained propaganda analysis was done by \citet{EMNLP19DaSanMartino}, who developed a corpus of news articles annotated with the spans of use of 18 propaganda techniques, from an invetory they put together. They targeted two tasks: (\emph{i})~binary classification ---given a sentence, predict whether any of the techniques was used in it; and (\emph{ii})~multi-label multi-class classification and span detection task ---given a raw text, identify both the specific text fragments where a propaganda technique is being used as well as the type of technique. They further proposed a multi-granular gated deep neural network that captures signals from the sentence-level task to improve the performance of the fragment-level classifier and vice versa. Subsequently, an automatic system, \texttt{Prta}, was developed and made publicly available \cite{da2020prta}, which performs fine-grained propaganda analysis of text using these 18 fine-grained propaganda techniques. 

\paragraph{Multimodal Content} Another line of related research is on analyzing multimodal content, e.g., for predicting misleading information \cite{Volkova_Ayton_Arendt_Huang_Hutchinson_2019}, for detecting deception \cite{Glenski2019MultilingualMD}, emotions and propaganda \cite{abd_kadir_etal}, hateful memes \cite{kiela2020hateful}, and propaganda in images \cite{doi:10.1080/15551393.2014.955501}. 
\citet{Volkova_Ayton_Arendt_Huang_Hutchinson_2019} developed a corpus of 500K Twitter posts consisting of images and labeled with six classes: disinformation, propaganda, hoaxes, conspiracies, clickbait, and satire. \citet{Glenski2019MultilingualMD} explored multilingual multimodal content for deception detection. Multimodal hateful memes were the target of the \emph{Hateful Memes Challenge}, which was addressed by fine-tuning state-of-art methods such as ViLBERT \cite{lu2019vilbert}, Multimodal Bitransformers \cite{kiela2019supervised}, and VisualBERT \cite{li2019visualbert} to classify hateful vs. not-hateful memes \cite{kiela2020hateful}.

\paragraph{Related Shared Tasks}
The present shared task is closely related to \emph{SemEval-2020 task 11 on Detection of Persuasion Techniques in News Articles} \cite{da-san-martino-etal-2020-semeval}, which focused on news articles, and asked (\emph{i})~to detect the spans where propaganda techniques are used, as well as (\emph{ii})~to predict which propaganda technique (from an inventory of 14 techniques) is used in a given text span. Another closely related shared task is the \emph{NLP4IF-2019 task on Fine-Grained Propaganda Detection}, which asked to detect the spans of use in news articles of each of 18 propaganda techniques \cite{da-san-martino-etal-2019-findings}. While these tasks focused on the text of news articles, here we target memes and multimodality, and we further use an extended inventory of 22 propaganda techniques.

Other related shared tasks include the FEVER 2018 and 2019 tasks on \emph{Fact Extraction and VERification}~\cite{thorne-EtAl:2018:N18-1}, 
the SemEval 2017 and 2019 tasks on predicting the veracity of rumors in Twitter \cite{derczynski-EtAl:2017:SemEval,gorrell-etal-2019-semeval}, the SemEval-2019 task on \emph{Fact-Checking in Community Question Answering Forums}~\cite{mihaylova-etal-2019-semeval}, the NLP4IF-2021 shared task on \emph{Fighting the COVID-19 Infodemic} \cite{NLP4IF-2021-COVID19-task}. We should also mention the CLEF 2018--2021 {\em CheckThat!} lab \cite{clef2018checkthat:overall,CheckThat:ECIR2019,clef-checkthat:2019,clef-checkthat:2020,CheckThat:ECIR2020}, which featured tasks on automatic identification \cite{clef2018checkthat:task1,clef-checkthat-T1:2019} and verification \cite{clef2018checkthat:task2,clef-checkthat-T2:2019,clef-checkthat-ar:2020,clef-checkthat-en:2020,10.1007/978-3-030-72240-1_75} of claims in political debates and social media. While these tasks focused on factuality, check-worthiness, and stance detection, here we target propaganda; moreover, we focus on memes and on multimodality rather than on analyzing the text of tweets, political debates, or community question answering forums.

%% file: sections/propaganda_techniques.tex
\section{Persuasion Techniques}
\label{sec:annotation}

Scholars have proposed a number of inventories of persuasion techniques of various sizes~\cite{Miller,Torok2015,article_kadir}. Here, we use an inventory of 22 techniques, borrowing from the lists of techniques described in \cite{EMNLP19DaSanMartino}, \cite{web_smear} and \cite{article_kadir}. Among these 22 techniques, the first 20 are applicable to both text and images, while the last two, \textit{Appeal to (Strong) Emotions} and \textit{Transfer}, are reserved for images. 

Below, we provide a definition for each of these 22 techniques; more detailed instructions of the annotation process and examples are provided in Appendix~\ref{sec:appendix_annotation_instructions}.

\begin{enumerate}[leftmargin=*]
   \itemsep0em
    \item \textbf{Loaded Language:} Using specific words and phrases with strong emotional implications (either positive or negative) to influence an audience.
    \item \textbf{Name Calling or Labeling:} Labeling the object of the propaganda campaign as either something the target audience fears, hates, finds undesirable, or loves, praises.
    \item \textbf{Doubt:} Questioning the credibility of someone or something.
    \item \textbf{Exaggeration or Minimisation:} Either representing something in an excessive manner, e.g.,~making things larger, better, worse (``\emph{the best of the best}'', ``\emph{quality guaranteed}''), or making something seem less important or smaller than it really is, e.g.,~saying that an insult was just a joke.
    \item \textbf{Appeal to Fear or Prejudices:} Seeking to build support for an idea by instilling anxiety and/or panic in the population towards an alternative. In some cases, the support is built based on preconceived judgments.
    \item \textbf{Slogans:} A brief and striking phrase that may include labeling and stereotyping. Slogans tend to act as emotional appeals.
    \item \textbf{Whataboutism:} A technique that attempts to discredit an opponent's position by charging them with hypocrisy without directly disproving their argument.
    \item \textbf{Flag-Waving:} Playing on strong national feeling (or positive feelings toward any group, e.g.,~based on race, gender, political preference) to justify or promote an action or idea.
    \item \textbf{Misrepresentation of Someone's Position (Straw Man):} When an opponent's proposition is substituted with a similar one, which is then refuted in place of the original proposition.
    \item \textbf{Causal Oversimplification:} Assuming a single cause or reason, when there are actually multiple causes for an issue. It includes transferring blame to one person or group of people without investigating the actual complexities of the issue.
    \item \textbf{Appeal to Authority:} Stating that a claim is true because a valid authority or expert on the issue said it was true, without any other supporting evidence offered. We consider the special case in which the reference is not an authority or an expert as part of this technique, although it is referred to as \emph{Testimonial} in the literature.
    \item \textbf{Thought-Terminating Clich\'{e}:} Words or phrases that discourage critical thought and meaningful discussion about a given topic. They are typically short, generic sentences that offer seemingly simple answers to complex questions or that distract the attention away from other lines of thought.
    \item \textbf{Black-and-White Fallacy or Dictatorship:} Presenting two alternative options as the only possibilities, when in fact more possibilities exist. As an extreme case, tell the audience exactly what actions to take, eliminating any other possible choices (\emph{Dictatorship}).
    \item \textbf{Reductio ad Hitlerum:} Persuading an audience to disapprove of an action or an idea by suggesting that the idea is popular with groups that are hated or in contempt by the target audience. It can refer to any person or concept with a negative connotation.
    \item \textbf{Repetition:} Repeating the same message over and over again, so that the audience will eventually accept it.
    \item \textbf{Obfuscation, Intentional Vagueness, Confusion:} Using words that are deliberately not clear, so that the audience can have their own interpretations.
    \item \textbf{Presenting Irrelevant Data (Red Herring):} Introducing irrelevant material to the issue being discussed, so that everyone's attention is diverted away from the points made.
    \item \textbf{Bandwagon} Attempting to persuade the target audience to join in and take the course of action because ``everyone else is taking the same action.''
    \item \textbf{Smears:} A smear is an effort to damage or call into question someone's reputation, by propounding negative propaganda. It can be applied to individuals or groups.
    \item \textbf{Glittering Generalities (Virtue):} These are words or symbols in the value system of the target audience that produce a positive image when attached to a person or an issue.
    \item \textbf{Appeal to (Strong) Emotions:} Using images with strong positive/negative emotional implications to influence an audience.
    \item \textbf{Transfer:} Also known as \emph{Association}, this is a technique that evokes an emotional response by projecting positive or negative qualities (praise or blame) of a person, entity, object, or value onto another one in order to make the latter more acceptable or to discredit it. 
\end{enumerate}

%% file: sections/dataset.tex
\section{Dataset}
\label{sec:dataset}

The annotation process is explained in detail in \mbox{Appendix \ref{sec:appendix_annotation_instructions}}, and in this section, we give a just brief summary.

We collected English memes from our personal Facebook accounts over several months in 2020 by following 26 public Facebook groups, which focus on politics, vaccines, COVID-19, and gender equality. We considered a meme to be a ``\emph{photograph style image with a
short text on top of it}'', and we removed examples that did not fit this definition, e.g.,~cartoon-style memes, memes whose textual content was strongly dominant or non-existent, memes with a single-color background image, etc.
Then, we annotated the memes using our 22 persuasion techniques. For each meme, we first annotated its textual content, and then the entire meme. We performed each of these two annotations in two phases: in the first phase, the annotators independently annotated the memes; afterwards, all annotators met together with a consolidator to discuss and to select the final gold label(s).

The final annotated dataset consists of 950 memes: 687 memes for training, 63 for development, and 200 for testing. While the maximum number of sentences in a meme is 13, the average number of sentences per meme is just 1.68, as most memes contain very little text.

Table~\ref{tab:instances_tasks} shows the number of instances of each technique for each of the tasks. Note that \textit{Transfer} and \textit{Appeal to (Strong) Emotions} are not applicable to text, i.e.,~to \mbox{Subtasks 1 and 2}. For \mbox{Subtasks 1 and 3}, each technique can be present at most once per example, while in \mbox{Subtask 2}, a technique could appear multiple times in the same example. This explains the sizeable differences in the number of instances for some persuasion techniques between \mbox{Subtasks 1 and 2}: some techniques are over-used in memes, with the aim of making the message more persuasive, and thus they contribute higher counts to Subtask 2.

\begin{table}[!tbh]
\centering
\setlength{\tabcolsep}{2.5pt}
\scalebox{0.75}{
 \begin{tabular}{@{}l@{}cccc@{}}
 \toprule
 \textbf{Persuasion Techniques}& \textbf{Subtask 1} &\multicolumn{2}{c}{\bf Subtask 2}  & \textbf{Subtask 3}  \\
  & \textbf{\#} &  \textbf{Len.} & \textbf{\#} & \textbf{\#}  \\
  \midrule
Loaded Language  & 489 & 2.41 & 761  & 492  \\
Name Calling/Labeling   & 300 & 2.62  & 408  &347  \\
Smears   & 263 & 17.11 & 266 & 602  \\
Doubt    & 84 & 13.71 &  86  &111 \\
Exaggeration/Minimisation  & 78  & 6.69 &  85  &  100\\
Slogans  & 66  & 4.70 &  72  & 70 \\
Appeal to Fear/Prejudice  & 57 & 10.12 &  60  & 91 \\
Whataboutism   & 54 & 22.83 &  54  & 67\\
Glittering Generalities (Virtue) & 44 & 14.07 &  45   &112  \\
Flag-Waving   & 38 & 5.18 &  44 &  55 \\
Repetition & 12 & 1.95 &  42  & 14   \\ 
Causal Oversimplification & 31 & 14.48 &  33   &  36 \\
Thought-Terminating Clich\'{e} & 27 & 4.07 &  28  & 27  \\
\begin{tabular}[c]{@{}l@{}}Black-and-White \\Fallacy/Dictatorship\end{tabular} & 25 & 11.92 &  25   & 26\\
Straw Man & 24 & 15.96 &  24  & 40 \\
Appeal to Authority & 22 & 20.05 &  22    & 35 \\
Reductio ad Hitlerum &  13 & 12.69 &  13  & 23  \\
\begin{tabular}[c]{@{}l@{}}Obfuscation, Intentional \\Vagueness, Confusion\end{tabular} & 5 &  9.8 &  5 & 7\\
Presenting Irrelevant Data & 5 &     15.4 &   5  & 7 \\
Bandwagon  & 5 &   8.4 &   5   & 5    \\
\hline
Transfer & --- & --- &  --- & 95  \\
Appeal to (Strong) Emotions & --- & --- & --- & 90  \\
\midrule
\textbf{Total} & \textbf{1,642} & & \textbf{2,119}  & \textbf{2,488} \\
 \bottomrule
\end{tabular}
}
\caption{Statistics about the persuasion techniques. For each technique, we show the average length of its spans (in number of words) and the number of its instances as annotated in the text only vs. in the entire meme.}
\label{tab:instances_tasks}
\end{table}

Note that the number of instances for Subtasks 1 and 3 differs, and in some cases by quite a bit, e.g.,~for \textit{Smears}, \textit{Doubt}, and \textit{Appeal to Fear/Prejudice}. This shows that many techniques cannot be found in the text, and require the visual content, which motivates the need for multimodal approaches for Subtask 3. Note also that different techniques have different span lengths, e.g., \emph{Loaded Language} and \emph{Name Calling} are about 2--3 words long, e.g.,~\textit{violence}, \textit{mass shooter}, and \textit{coward}. However, for techniques such as \textit{Whataboutism}, the average span length is 22 words.

\begin{figure}[!htp]
\centering
\subfloat[Subtask 1]{\label{stsk1}%
  \includegraphics[width=\columnwidth]{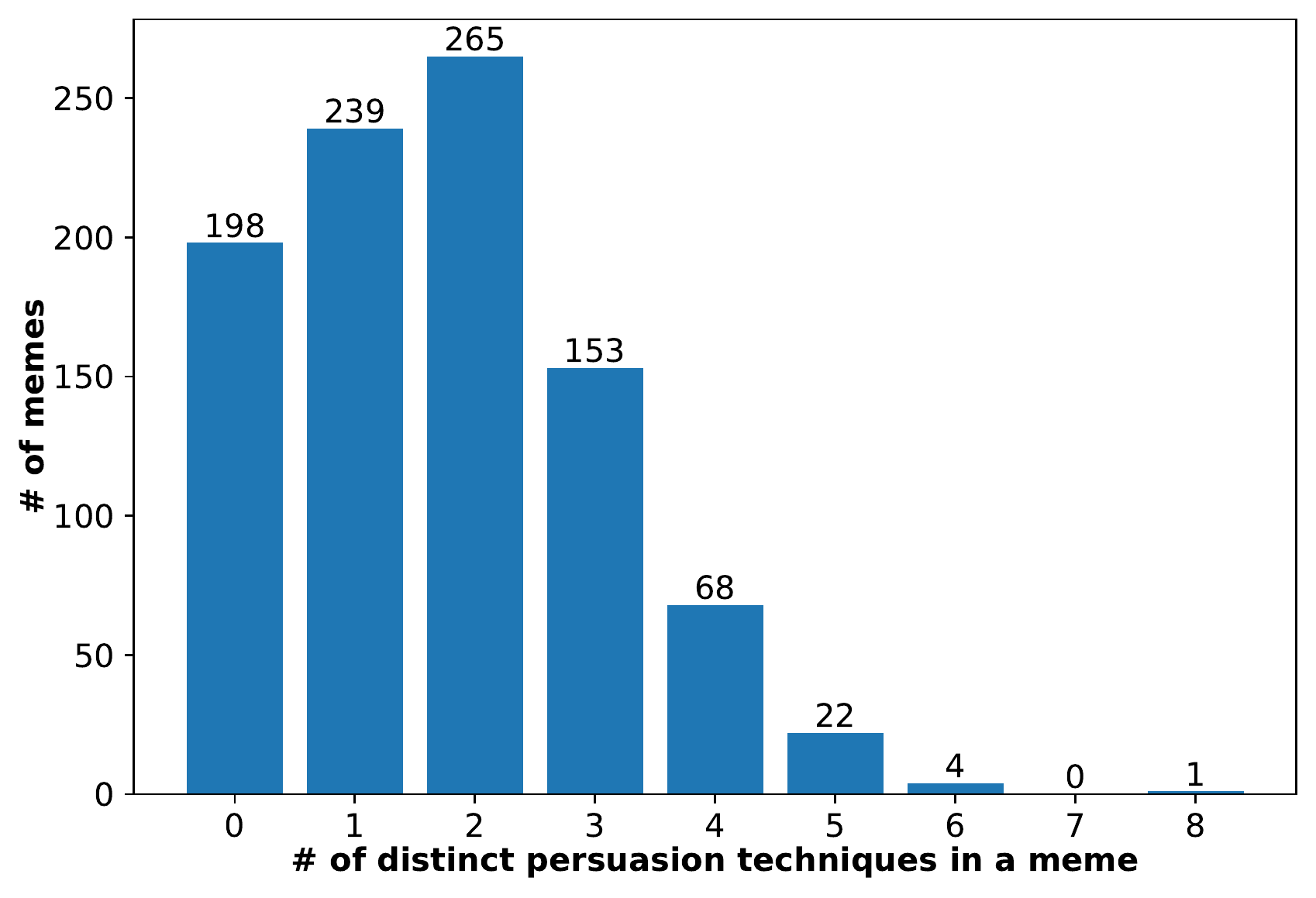}%
}
\vspace{0.5cm}
\subfloat[Subtask 2]{\label{stsk2}%
  \includegraphics[width=\columnwidth]{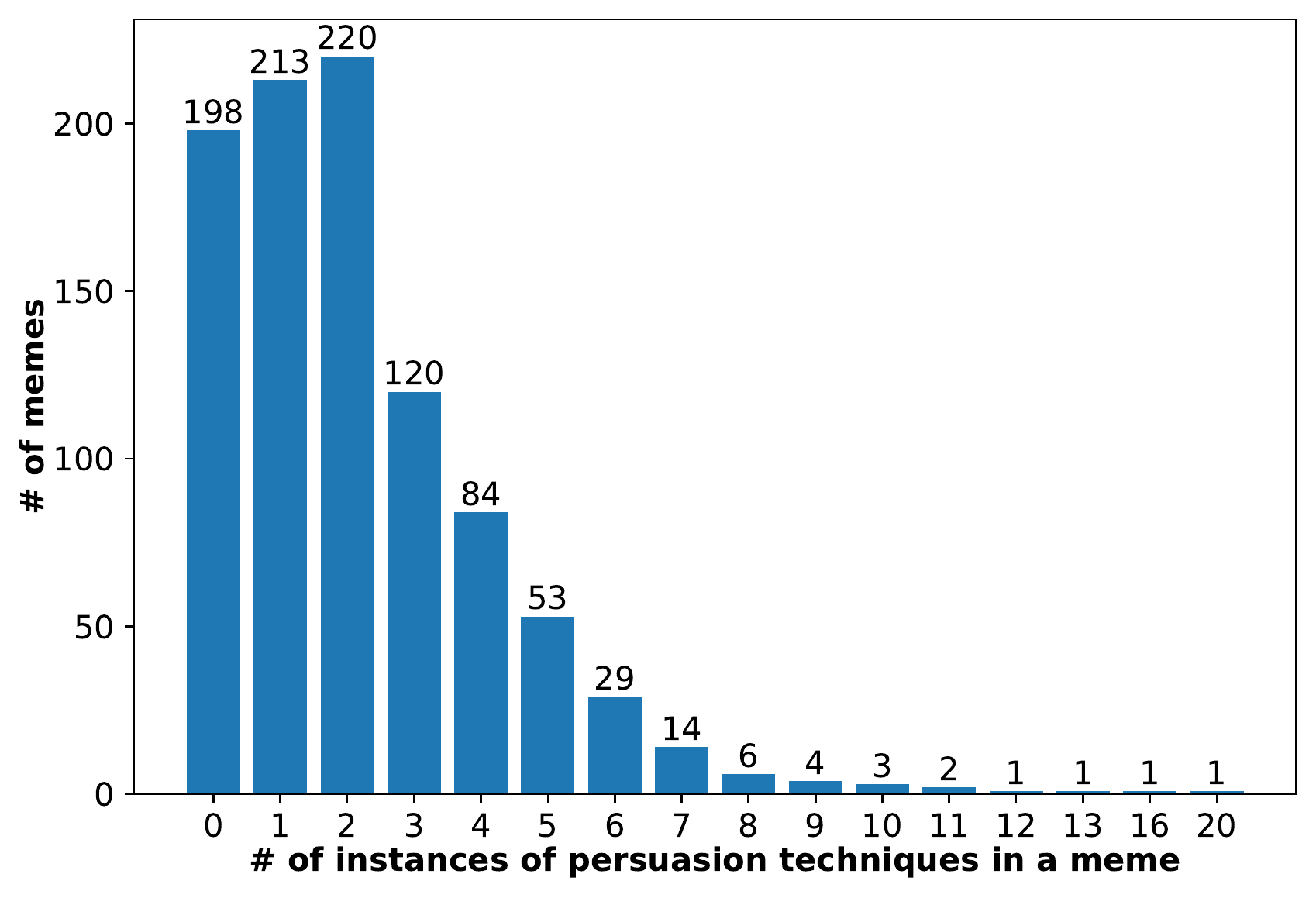}%
}
\vspace{0.5cm}
\subfloat[Subtask 3]{\label{stsk3}%
  \includegraphics[width=\columnwidth]{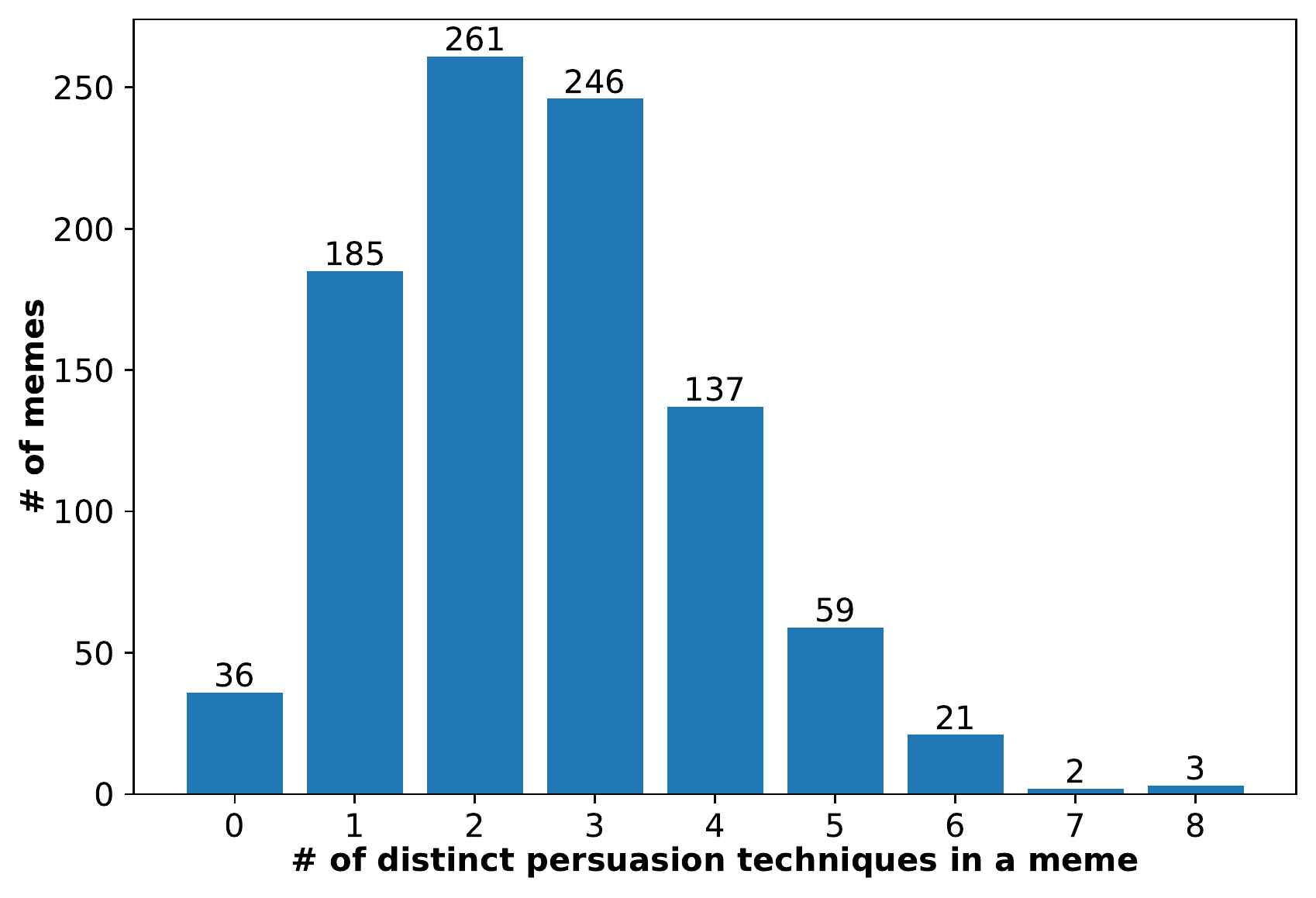}%
}
\caption{Distribution of the number of persuasion techniques per meme. Subfigure (b) reports the \textbf{number of instances} of persuasion techniques for a meme. Note that a meme could have multiple instances of the same technique for this subtask. Subfigures (a) and (c) show the number of \textbf{distinct} persuasion techniques in a meme.}
\label{fig:label_disttext}
\end{figure}

Figure~\ref{fig:label_disttext} shows statistics about the distribution of the number of persuasion techniques per meme.
Note the difference for memes without persuasion techniques between Figures~\ref{stsk1} and \ref{stsk3}: we can see that the number of memes without any persuasion technique drastically drops for \mbox{Subtask 3}. This is because the visual modality introduces additional context that was not available during the text-only annotation, which further supports the need for multimodal analysis. The visual modality also has an impact on memes that already had persuasion techniques in the text-only phase. 

We observe that the number of memes with only one persuasion technique in \mbox{Subtask 3} is considerably lower compared to \mbox{Subtask 1}, while the number of memes with three or more persuasion techniques has greatly increased for \mbox{Subtask 3}.

%% file: sections/evaluation_framework.tex
\section{Evaluation Framework}
\label{sec:evaluation_framework}

\subsection{Evaluation Measures}
\label{ssec:evaluation_measures}

\paragraph{Subtasks 1 and 3}
To measure the performance of the systems, for Subtasks 1 and 3, we use Micro and Macro F$_1$, as these are multi-class multi-label tasks, where the labels are imbalanced. 
The official measure for the task is Micro F$_1$. 

\paragraph{Subtask 2}
For Subtask 2, the evaluation requires matching the text spans. Hence, we use an evaluation function that gives credit to partial matches between gold and predicted spans.

\begin{figure}[h]
    \centering
    \scalebox{0.9}{
    \begin{tikzpicture}
        \tikzstyle{box} = [draw, fill=white, row sep=0.1cm,column sep=0.01cm, text width=0.2cm, minimum height=0.5cm, text height=0.17cm, rectangle, align=center, inner sep=0.04cm,outer sep=0.0cm];
        \def\goldsentence{h,o,w, ,s,t,u,p,i,d, ,a,n,d, ,p,e,t,t,y, ,t,h,i,n,g,s} 
        \foreach[count=\xi] \x in \goldsentence \node[box] (t\xi) at (0.28*\xi,0) {\textbf{\scriptsize \x}};
        \draw [decorate,decoration={brace,amplitude=4pt}] (1.3,0.30) -- (5.7,0.30) node [black,midway,yshift=0.3cm] {\footnotesize $t_1$: loaded language};
        \foreach[count=\xi] \x in \goldsentence \node[box] (s\xi) at (0.28*\xi,-1.25) {\textbf{\scriptsize \x}};
        \draw [decorate,decoration={brace,amplitude=4pt}] (1.0,-0.95) -- (2.9,-0.95) node [black,midway,yshift=0.3cm] {\footnotesize $s_1$: loaded language};
        \draw [decorate,decoration={brace,amplitude=4pt}] (4.3,-0.95) -- (5.7,-0.95) node [black,midway,yshift=0.3cm] {\footnotesize $s_2$: name calling};
        \foreach[count=\xi] \x in \goldsentence \node[box] (s\xi) at (0.28*\xi,-2.5) {\textbf{\scriptsize \x}};
        \draw [decorate,decoration={brace,amplitude=4pt}] (1.3,-2.2) -- (2.9,-2.2) node [black,midway,yshift=0.3cm] {\footnotesize $s_3$: loaded language};
        \draw [decorate,decoration={brace,amplitude=4pt}] (4.3,-2.2) -- (5.7,-2.2) node [black,midway,yshift=0.3cm] {\footnotesize $s_5$: loaded language};
        \draw [decorate,decoration={mirror,brace,amplitude=4pt}] (3.22,-2.8) -- (4.05,-2.8) node [black,midway,yshift=-0.35cm] {\footnotesize $s_4$: loaded language};
    \end{tikzpicture}
    }
    \caption{Example of gold annotation (top) and the predictions of a supervised model (bottom) in a document represented as a sequence of characters.}
    \label{fig:metrics_example}
\end{figure}
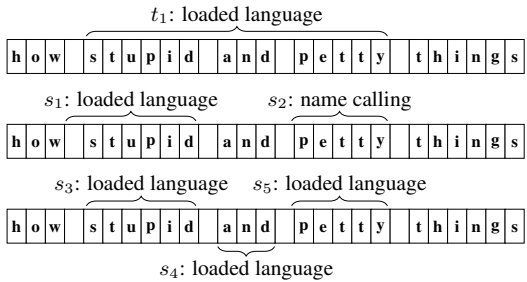

Let document $\boldsymbol{d}$ be represented as a sequence of characters. 
The $i$-th propagandistic text fragment is then represented as a sequence of contiguous characters $t\subseteq \boldsymbol{d}$. 
A document includes a set of (possibly overlapping) fragments $T$. 
Similarly, a learning algorithm produces a set $S$ with fragments $s\subseteq \boldsymbol{d}$, predicted on $\boldsymbol{d}$. 
A labeling function $l(x) \in \{1,\ldots,20\}$ associates $t\in T$, $s\in S$ with one of the techniques. 
An example of (gold) annotation is shown in Figure~\ref{fig:metrics_example}, where an annotation $t_1$ marks the span \emph{stupid and petty} with the technique \emph{Loaded Language}.

We define the following function to handle partial overlaps of fragments with the same labels:
\begin{equation}
    C(s,t,h) = \frac{|(s\cap t)|}{h}\delta\left(l(s), l(t) \right), 
    \label{eq:annotationsScore}
\end{equation}
where $h$ is a normalizing factor and $\delta(a,b)=1$ if $a=b$, and 0, otherwise. 
For example, still with reference to Figure~\ref{fig:metrics_example}, $C(t_1,s_1,|t_1|)=\frac{6}{16}$ and $C(t_1,s_2,|t_1|) = 0$.

Given Eq. \eqref{eq:annotationsScore}, we now define variants of precision and recall that can account for the imbalance in the corpus:
\begin{equation}
P(S,T) = \frac{1}{|S|}\!\sum_{\begin{minipage}[t]{0.97cm}\footnotesize $s\in S,\\$ $t\in T$\end{minipage}}\!\!\! C(s,t,|s|), 
\label{eq:precisiontask3}
\end{equation}
\begin{equation}
R(S,T) = \frac{1}{|T|}\!\sum_{\begin{minipage}[t]{0.97cm}\footnotesize $s\in S,\\$ $t\in T$\end{minipage}}\!\!\! C(s,t,|t|),
\label{eq:recalltask3}
\end{equation}

We define \eqref{eq:precisiontask3} to be zero if $|S|=0$, and Eq.~\eqref{eq:recalltask3} to be zero if $|T|=0$. 
Following~\citet{Potthast2010a}, in \eqref{eq:precisiontask3} and~\eqref{eq:recalltask3} we penalize systems predicting too many or too few instances by dividing by $|S|$ and $|T|$, respectively.
Finally, we combine Eqs.~\eqref{eq:precisiontask3} and \eqref{eq:recalltask3} into an F$_1$-measure, the harmonic mean of precision and recall.

\subsection{Task Organization}
\label{task_organization}

We ran the shared task in two phases:

\paragraph{Development Phase} In the first phase, only training and development data were made available, and no gold labels were provided for the latter. The participants competed against each other to achieve the best performance on the development set. A live 
leaderboard was made available to keep track of all submissions.

\paragraph{Test Phase} In the second phase, the test set was released and the participants were given just a few days to submit their final predictions. 

\paragraph{}In the \textit{Development Phase}, the participants could make an unlimited number of submissions, and see the outcome in their private space. The best score for each team, regardless of the submission time, was also shown in a public leaderboard. As a result, not only could the participants observe the impact of various modifications in their systems, but they could also compare against the results by other participating teams. 
In the \textit{Test Phase}, the participants could again submit multiple runs, but they would not get any feedback on their performance. Only the latest submission of each team was considered as official and was used for the final team ranking. 
The final leaderboard on the test set was made public after the end of the shared task. 

In the \textit{Development Phase}, a total of 15, 10 and 13 teams made at least one submission for ST1, ST2 and ST3, respectively. 
In the \textit{Test Phase} the number of teams who made official submissions was 16, 8, and 15 for ST1, ST2, ST3, respectively. 


After the competition was over, we left the submission system open for the development set, and we plan to reopen it on the test set as well. The up-to-date leaderboards can be found on the website of the competition.\footnote{\url{http://propaganda.math.unipd.it/semeval2021task6/}}

\begin{table*}[tbh]
\centering
\setlength{\tabcolsep}{1.7pt}    
\footnotesize
\scalebox{1.0}{
\begin{tabular}{l|llllll|llllll|lll|lll}
\toprule
\multicolumn{1}{c}{\textbf{Rank. Team}} & \multicolumn{6}{c}{\textbf{Transformers}} & \multicolumn{6}{c}{\textbf{Models}} & \multicolumn{3}{c}{\textbf{Repres.}} & \multicolumn{3}{c}{\textbf{Misc}} \\\midrule
  & \rotatebox{90}{\textbf{BERT}} & \rotatebox{90}{\textbf{RoBERTa}} & \rotatebox{90}{\textbf{XLNet}} & \rotatebox{90}{\textbf{ALBERT}} & \rotatebox{90}{\textbf{DistilBERT}} & \rotatebox{90}{\textbf{DeBERTa}} & \rotatebox{90}{\textbf{LSTM}} & \rotatebox{90}{\textbf{CNN}} & \rotatebox{90}{\textbf{SVM}} & \rotatebox{90}{\textbf{Naive Bayes}} & \rotatebox{90}{\textbf{Random Forest}} & \rotatebox{90}{\textbf{CRF}} & \rotatebox{90}{\textbf{Embeddings}} & \rotatebox{90}{\textbf{Char n-grams}} & \rotatebox{90}{\textbf{PoS}} & \rotatebox{90}{\textbf{Ensemble}} & \rotatebox{90}{\textbf{Data augmentation}} & \rotatebox{90}{\textbf{Postprocessing}} \\ \midrule
1. MinD & \sq & \sq & \sq & \sq &  & \sq & & & & & & & \sq & \sq & & \sq & & \sq \\
2. Alpha & & & & & & \sq & & & & & & & & &  &  &  &  \\
3. Volta & \cq & \sq & & & & & & & & & & & \sq & & & & & \\
5. AIMH & \sq & & & & & &  & & & & &  & \sq & & &  & & \\
6. LeCun & & \cq & & & & \cq & \sq & & & & & & \sq & & & & \cq & \\
7. WVOQ & & & & & & & & & & & & & & & \sq & & & \\
9. NLyticsFKIE & & \sq & & &  & & & & & & & \sq & \cq & & & & \sq & \\
12. YNU-HPCC & \sq & \sq & & \sq &  & & & \sq & & & & & \sq & & & & & \\
13. CSECUDSG & & & & & \sq & & & & & & & & & & & \sq & & \\
15. NLP-IITR & \cq & \sq & & & & & \cq & & \cq & \cq & \cq & & \sq & & & & \sq & \\\bottomrule
\end{tabular}
}

\setlength{\tabcolsep}{1.2pt}
\begin{tabular}{@{}rl@{}}
1 & \cite{SemEval2021-6-Tian} \\
2 & \cite{SemEval2021-6-alpha} \\
3 & \cite{SemEval2021-6-Volta} \\
5 & \cite{SemEval2021-6-AIMH}
\end{tabular}
\hspace{0mm}
\begin{tabular}{@{}rl@{}}
6 & \cite{SemEval2021-6-LeCun}  \\
7 & \cite{SemEval2021-6-Roele} \\
9 & \cite{SemEval2021-6-NLyticsFKIE} \\
12 & \cite{SemEval2021-6-YNUHPCC} 
\end{tabular}
\hspace{0mm}
\begin{tabular}{@{}rl@{}}
13 & \cite{SemEval2021-6-Hossain} \\
15 & \cite{SemEval2021-6-VGupta} \\
\\
\\
\end{tabular}

\caption{\textbf{ST1:} Overview of the approaches used by the participating systems. \sq$=$part of the official submission; \cq$=$considered in internal experiments; \emph{Repres.} stand for Representations. References to system description papers are shown below the table.
}
\label{tab:overview_task1}
\end{table*}

%% file: sections/participants_systems.tex
\section{Participants and Results}
\label{sec:participants_systems}

Below, we give a general description of the systems that participated in the three subtasks and their results, with focus on those ranked among the \mbox{top-3}. 
Appendix~\ref{app:appendix_summary_sub_systems} gives a description of every system.

\subsection{Subtask 1 (Unimodal: Text)}

Table \ref{tab:overview_task1} gives an overview of the systems that took part in Subtask 1. We can see that transformers were quite popular, and among them, most commonly used was RoBERTa, followed by BERT. Some participants used learning models such as LSTM, CNN, and CRF in their final systems, while internally, Na\"{i}ve Bayes and Random Forest were also tried. In terms of representation, embeddings clearly dominated. Moreover, techniques such as ensembles, data augmentation, and post-processing were also used in some systems.

\begin{table}[h!]
\centering
\setlength{\tabcolsep}{2pt}
\resizebox{\linewidth}{!}{
\begin{tabular}{clcc}
	\toprule 
	\bf Rank & \bf Team & \bf F1-Micro & \bf F1-Macro \\
	\midrule
	1 & MinD & \bf .593 & .290$_{\textsc 2~}$   \\
	2 & Alpha & .572 & .262$_{\textsc 5~}$  \\
	3 & Volta & .570 & .266$_{\textsc 3~}$  \\
	4 & mmm & .548 & \bf .303$_{\textsc 1~}$  \\
	5 & AIMH & .539 & .245$_{\textsc 6~}$  \\
	6 & LeCun & .512 & .227$_{\textsc 8~}$  \\
	7 & WVOQ & .511 & .227$_{\textsc 8~}$  \\
	8 & TeamUNCC & .510 & .236$_{\textsc 7~}$  \\
	9 & NLyticsFKIE & .498 & .140$_{\textsc 13}$  \\
	10 & TeiAS & .497 & .214$_{\textsc 10}$  \\
	11 & DAJUST & .497 & .187$_{\textsc 11}$  \\
	12 & YNUHPCC & .493 & .263$_{\textsc 4~}$  \\
	13 & CSECUDSG & .489 & .185$_{\textsc 12}$  \\
	14 & TeamFPAI & .406 & .115$_{\textsc 15}$  \\
	15 & NLPIITR & .379 & .126$_{\textsc 14}$  \\
	   & \it Majority baseline &  \it .374 & \it .033 \\
	16 & TriHeadAttention & .184 & .024$_{\textsc 18}$  \\
	   & \it Random baseline & \it .064 & \it .044  \\
	\bottomrule
\end{tabular}
}
\caption{Results for Subtask 1. The systems are ordered by the official score: \emph{F1-micro}.}
\label{tab:result_task_1}
\end{table}

The evaluation results are shown in Table~\ref{tab:result_task_1}, which also includes two baselines: (\emph{i})~random, and (\emph{ii})~majority class. The latter always predicts \emph{Loaded Language}, as it is the most frequent technique for Subtask 1 (see Table \ref{tab:instances_tasks}). 

The best system \textbf{MinD}~\cite{SemEval2021-6-Tian} used five transformers: BERT, RoBERTa, XLNet, DeBERTa, and ALBERT. It was fine-tuned on the PTC corpus~\cite{da-san-martino-etal-2020-semeval} and then on the training data for \mbox{Subtask 1}. 

The final prediction for MinD averages the probabilities for these models, and further uses post-processing rules, e.g.,~each bigram appearing more than three times is flagged as a \textit{Repetition}.

Team \textbf{Alpha}~\cite{SemEval2021-6-alpha} was ranked second. However, they used features from images, which was not allowed (images were only allowed for Subtask 3).

Team \textbf{Volta}~\cite{SemEval2021-6-Volta} was third. They used a combination of transformers with the [CLS] token as an input to a two-layer feed-forward network. They further used example weighting to address class imbalance.

We should also mention team \textbf{LeCun}, which used additional corpora such as the PTC corpus \citep{da-san-martino-etal-2020-semeval}, and augmented the training data using synonyms, random insertion/deletion, random swapping, and back-translation.

\subsection{Subtask 2 (Unimodal: Text)}

The approaches for this task varied from modeling it as a question answering (QA) task to performing multi-task learning. 
Table \ref{tab:overview_task2} presents a high-level summary. We can see that BERT dominated, while RoBERTa was much less popular. We further see a couple of systems using data augmentation. Unfortunately, there are too few systems with system description papers for this subtask, and thus it is hard to do a very deep analysis.

\begin{table}[h!]
\centering
\setlength{\tabcolsep}{1.2pt}    
\footnotesize
\scalebox{1.0}{
\begin{tabular}{@{}l|ll|lll|llll|ll@{}}
\toprule
\multicolumn{1}{c}{\textbf{Rank. Team}} & \multicolumn{2}{c}{\textbf{Trans.}} & \multicolumn{3}{c}{\textbf{Models}} & \multicolumn{4}{c}{\textbf{Repres.}} & \multicolumn{2}{c}{\textbf{Misc}} \\ \midrule 
 & \rotatebox{90}{\textbf{BERT}} & \rotatebox{90}{\textbf{RoBERTa}} & \rotatebox{90}{\textbf{LSTM}} & \rotatebox{90}{\textbf{CNN}} & \rotatebox{90}{\textbf{SVM}} & \rotatebox{90}{\textbf{ELMo}} & \rotatebox{90}{\textbf{PoS}} & \rotatebox{90}{\textbf{Sentiment}} & \rotatebox{90}{\textbf{Rhetorics}} & \rotatebox{90}{\textbf{Ensemble}} & \rotatebox{90}{\textbf{Data augmentation}} \\ \midrule
1. Volta & \cq & \sq & & & & & & & & & \\
2. HOMADOS & \sq & & & & &  & & & & & \\
3. TeamFPAI & \sq & & & & & & & & & \sq & \sq \\
5. WVOQ & \sq & & \sq & & \cq & & \sq & \sq & \sq & & \\
6. CSECUDSG & \sq & & & & & & \cq & & & & \sq \\
7. YNU-HPCC & \sq & & & & & & & & & & \\ \bottomrule
\end{tabular}
}

\setlength{\tabcolsep}{1.2pt}
\begin{tabular}{@{}rl@{}}
1 & \cite{SemEval2021-6-Volta} \\
2 & \cite{SemEval2021-6-Kaczynski} \\
3 & \cite{SemEval2021-6-Hou}
\end{tabular}
\hspace{2mm}
\begin{tabular}{@{}rl@{}}
5 & \cite{SemEval2021-6-Roele} \\
6 & \cite{SemEval2021-6-Hossain} \\
7 & \cite{SemEval2021-6-YNUHPCC}
\end{tabular}
\caption{\textbf{ST2:} Overview of the approaches used by the participating systems. \sq$=$part of the official submission; \cq$=$considered in internal experiments; \emph{Trans.} is for Transformers; \emph{Repres.} is for Representations. References to system description papers are shown below the table.
}
\label{tab:overview_task2}
\end{table}

\begin{table}[tbh]
\centering
\setlength{\tabcolsep}{2.5pt}
\resizebox{\linewidth}{!}{
\begin{tabular}{clccc}
	\toprule 
	\bf Rank & \bf Team & \bf F1 & \bf Precision & \bf Recall \\
	\midrule
	1 & Volta  & \bf \bf .482 & .501$_{\textsc 2}$ & \bf .464$_{\textsc 1}$ \\
	2 & HOMADOS & .407 & .412$_{\textsc 3}$ & .403$_{\textsc 2}$ \\
	3 & TeamFPAI & .397 & \bf .652$_{\textsc 1}$ & .286$_{\textsc 5}$ \\
	4 & TeamUNCC & .329 & .285$_{\textsc 4}$ & .390$_{\textsc 3}$ \\
	5 & WVOQ & .268 & .243$_{\textsc 5}$ & .299$_{\textsc 4}$ \\
	6 & CSECUDSG & .120 & .080$_{\textsc 8}$ & .243$_{\textsc 6}$ \\
	7 & YNUHPCC & .091 & .186$_{\textsc 6}$ & .060$_{\textsc 7}$ \\
	8 & TriHeadAttention & .080 & .170$_{\textsc 7}$ & .052$_{\textsc 8}$ \\
	  & \it Random Baseline & \it .010 & \it .034 & \it .006 \\
	\bottomrule
\end{tabular}
}
\caption{Results for Subtask 2. The systems are ordered by the official score: \emph{F1-micro}.}
\label{tab:result_task_2}
\end{table}

Table~\ref{tab:result_task_2} shows the evaluation results. We report our random baseline, which is based on the random selection of spans with random lengths and a random assignment of labels.

The best model by team \textbf{Volta}~\cite{SemEval2021-6-Volta} used various transformer models, such as BERT and RoBERTa, to predict token classes by considering the output of each token embedding. Then, they assigned classes for a given word as the union of the classes predicted for the subwords that make that word (to account for BPEs).

Team \textbf{HOMADOS}~\cite{SemEval2021-6-Kaczynski} was second, and they used a multi-task learning (MTL) and additional datasets such as the PTC corpus from SemEval-2020 task 11~\cite{da-san-martino-etal-2020-semeval}, and a fake news corpus \cite{Przybyla_2020}. They used BERT, followed by several output layers that perform auxiliary tasks of propaganda detection and credibility assessment in two distinct scenarios: sequential and parallel MTL. Their final submission used the latter.

Team \textbf{TeamFPAI~\cite{SemEval2021-6-Hou}} formulated the task as a question answering problem using machine reading comprehension, thus improving over the ensemble-based approach of \citet{liu-etal-2018-multi}. They further explored data augmentation and loss design techniques, in order to alleviate the problem of data sparseness and data imbalance.

\subsection{Subtask 3 (Multimodal: Memes)}

\begin{table*}[!hbt]
\centering
\setlength{\tabcolsep}{1.2pt}    
\footnotesize
\scalebox{0.9}{
\begin{tabular}{@{}l|lllllll|llllllll|llllllllllllll|lllll|lll@{}}
\toprule
\multicolumn{1}{c}{\textbf{Rank. Team}} & \multicolumn{7}{c}{\textbf{Transformers}} & \multicolumn{8}{c}{\textbf{Models}} & \multicolumn{14}{c}{\textbf{Representations}} & \multicolumn{5}{c}{\textbf{Fusion}} & \multicolumn{3}{c}{\textbf{Misc}} \\ \midrule
 & \rotatebox{90}{\textbf{BERT}} & \rotatebox{90}{\textbf{RoBERTa}} & \rotatebox{90}{\textbf{XLNet}} & \rotatebox{90}{\textbf{ALBERT}} & \rotatebox{90}{\textbf{FastBERT}} & \rotatebox{90}{\textbf{GPT-2}} & \rotatebox{90}{\textbf{DeBERTa}} & \rotatebox{90}{\textbf{ResNet18}} & \rotatebox{90}{\textbf{ResNet50}} & \rotatebox{90}{\textbf{ResNet51}} & \rotatebox{90}{\textbf{VGG16}} & \rotatebox{90}{\textbf{LSTM}} & \rotatebox{90}{\textbf{CNN}} & \rotatebox{90}{\textbf{SVM}} & \rotatebox{90}{\textbf{CRF}} & \rotatebox{90}{\textbf{Embeddings}} & \rotatebox{90}{\textbf{ELMo}} & \rotatebox{90}{\textbf{Words/Word n-grams}} & \rotatebox{90}{\textbf{Char n-grams}} & \rotatebox{90}{\textbf{PoS}} & \rotatebox{90}{\textbf{Sentiment}} & \rotatebox{90}{\textbf{Rhetorics}} & \rotatebox{90}{\textbf{FR (ResNet34)}} & \rotatebox{90}{\textbf{MS OCR}} & \rotatebox{90}{\textbf{YouTube-8M}} & \rotatebox{90}{\textbf{CLIP}} & \rotatebox{90}{\textbf{BUTD}} & \rotatebox{90}{\textbf{ERNIE-VIL}} & \rotatebox{90}{\textbf{SemVLP}} & \rotatebox{90}{\textbf{Average}} & \rotatebox{90}{\textbf{Concat}} & \rotatebox{90}{\textbf{Attention}} & \rotatebox{90}{\textbf{MLP}} & \rotatebox{90}{\textbf{Chained classifier}} & \rotatebox{90}{\textbf{Ensemble}} & \rotatebox{90}{\textbf{Data augmentation}} & \rotatebox{90}{\textbf{Postprocessing}} \\ \midrule
1. Alpha & \cq & \sq & \cq & &  & \cq & &  & \cq &   &  & & & & & \sq & & & & & & & & &  & & \cq & \cq & & & & & & & \cq & & \\
2. MinD & \sq & \sq & \sq & \sq &  & & \sq &  &  &   &  & & \sq & & \sq & \sq & & & \sq & & & & \sq & \sq &  & &  &  & \sq & \sq & \cq & & \cq & & \sq & & \sq \\
3. 1213Li & & \sq & & & & & & & \sq &  & & & & & & & & & & & & & & &  & &  &  & & \sq & & \sq & & & & & \\
4. AIMH & \sq & & & &  & & &  & \sq & & &  & & &  &  &  & & & & & & & &  & &  &  & & & \sq & & \sq & &  & & \\
5. Volta & \cq & \sq & & &  & & &  &  &   &  & & \sq & & & \sq & & & & & & & & &  & &  &  & & & \sq & & & & \sq & & \\
6. CSECUDSG & \sq & \sq & & & \sq & & & & \sq & & & & & & & \sq & & & & \cq & & & & & \sq & &  &  & & & \sq & & \sq & & \sq & & \\
8. LIIR & \sq & & & &  & & &  &  &   &  & & & & & & & \sq & & & & & & &  & \sq &  &  & & & & & & \sq & & \sq & \\
10. WVOQ & \sq & & & &  & & &  &  & &    & \sq & & \cq & & & & & & \sq & \sq & \sq & & &  & &  &  & & & & & & & & & \\
11. YNU-HPCC & & & & \sq &  & & & \sq &  &  & \sq & & \sq & & & & & & & & & & & &  & &  &  & & & \sq & & \sq & & & & \\
13. NLyticsFKIE & \sq & & & &  & & &  &  &   &  & \cq & & & \sq & \cq & & & & & & & & &  & &  &  & & & & & & & & \sq & \\
15. LT3-UGent & \sq & & & &  & & &  &  & \sq   &  & & & & \sq & & & & & & & & & &  & &  &  & & & & & & & & \sq & \\ \bottomrule
\end{tabular}
}

\setlength{\tabcolsep}{1.2pt} 
\begin{tabular}{@{}rl@{}}
1 & \cite{SemEval2021-6-alpha} \\
2 & \cite{SemEval2021-6-Tian} \\
3 & \cite{SemEval2021-6-Li} \\
4 & \cite{SemEval2021-6-AIMH} \\
\end{tabular}
\hspace{2mm}
\begin{tabular}{@{}rl@{}}
5 & \cite{SemEval2021-6-Volta} \\
6 & \cite{SemEval2021-6-Hossain}\\
8 & \cite{SemEval2021-6-Ghadery} \\
10 & \cite{SemEval2021-6-Roele} \\
\end{tabular}
\hspace{2mm}
\begin{tabular}{@{}rl@{}}
11 & \cite{SemEval2021-6-YNUHPCC} \\
13 & \cite{SemEval2021-6-NLyticsFKIE} \\
15 & \cite{SemEval2021-6-Singh}\\
\\
\end{tabular}
\caption{\textbf{ST3:} Overview of the approaches used by the participating systems. \sq$=$part of the official submission; \cq$=$considered in internal experiments. References to system description papers are shown below the table.}
\label{tab:overview_task3}
\end{table*}

Table \ref{tab:overview_task3} presents an overview of the approaches used by the systems that participated in Subtask 3. This is a very rich and very interesting table. We can see that transformers were quite popular for text representation, with BERT dominating, but RoBERTa being quite popular as well. For the visual modality, the most common representations were variants of ResNet, but VGG16 and CNNs were also used. We further see a variety of representations and fusion methods, which is to be expected given the multi-modal nature of this subtask.

Table \ref{tab:result_task_3} shows the performance on the test set for the participating systems for Subtask 3. The two baselines shown in the table are similar to those for Subtask 1, namely a random baseline and a majority class baseline. However, this time the most frequent class baseline always predicts \emph{Smears} (for Subtask 1, it was \emph{Loaded Language}), as this is the most frequent technique for Subtask 3 (as can be seen in Table \ref{tab:instances_tasks}).

Team \textbf{Alpha}~\cite{SemEval2021-6-alpha} pre-trained a transformer using text with visual features. They extracted grid features using ResNet50, and salient region features using BUTD. They further used these grid features to capture the high-level semantic information in the images. Moreover, they used salient region features to describe objects and to caption the event present in the memes. Finally, they built an ensemble of fine-tuned DeBERTA+ResNet, DeBERTA+BUTD, and ERNIE-VIL systems.

Team \textbf{MinD} \cite{SemEval2021-6-Tian} combined a system for Subtask 1 with (\emph{i})~ResNet-34, a face recognition system, (\emph{ii})~OCR-based positional embeddings for text boxes, and (\emph{iii})~Faster R-CNN to extract region-based image features. They used late fusion to combine the textual and the visual representations.
Other multimodal fusion strategies they tried were concatenation of the representation and mapping using a multi-layer perceptron.

Team \textbf{1213Li}~\cite{SemEval2021-6-Li} used RoBERTa and ResNet-50 as feature extractors for texts and images, respectively, and adopted a label embedding layer with a multi-modal attention mechanism to measure the similarity between labels with multi-modal information, and fused features for label prediction.

\begin{table}[h!]
\centering
\setlength{\tabcolsep}{2.5pt}
\resizebox{\linewidth}{!}{
\begin{tabular}{@{}llcc@{}}
\toprule
\multicolumn{1}{l}{\textbf{Rank}} & \multicolumn{1}{l}{\textbf{Team}} & \multicolumn{1}{c}{\textbf{F1-Micro}} & \multicolumn{1}{c}{\textbf{F1-Macro}} \\ \midrule
1 & Alpha & \textbf{.581} & \textbf{.273}$_{\textsc 1~}$ \\
2 & MinD & .566 & .244$_{\textsc 3~}$ \\
3 & 1213Li & .549 & .228$_{\textsc 5~}$ \\
4 & AIMH & .540 & .207$_{\textsc 6~}$ \\
5 & Volta & .521 & .189$_{\textsc 8~}$ \\
6 & CSECUDSG & .513 & .121$_{\textsc 11}$ \\
7 & aircasMM & .511 & .200$_{\textsc 7~}$ \\
8 & LIIR & .498 & .188$_{\textsc 9~}$ \\
9 & CAU731NLP & .481 & .084$_{\textsc 14}$ \\
10 & WVOQ & .478 & .240$_{\textsc 4~}$ \\
11 & YNUHPCC & .446 & .096$_{\textsc 13}$ \\
12 & TriHeadAttention & .442 & .062$_{\textsc 15}$ \\
13 & NLyticsFKIE & .423 & .118$_{\textsc 12}$ \\
 & \it Majority baseline & \it .354 & \it .036~~ \\
14 & LT3UGent & .332 & .264$_{\textsc 2~}$ \\
15 & TeamUNCC & .224 & .124$_{\textsc 10}$ \\
 & \it Random baseline & \it .071 & \it .052~~ \\ \bottomrule
\end{tabular}
}
\caption{Results for Subtask 3. The systems are ordered by the official score: \emph{F1-micro}.}
\label{tab:result_task_3}
\end{table}

%% file: sections/conclusion.tex
\section{Conclusion and Future Work}
\label{sec:conclusions}

We presented \emph{SemEval-2021 Task 6 on Detection of Persuasion Techniques in Texts and Images}. It was a successful task: a total of 71 teams registered to participate, 22 teams eventually made an official submission on the test set, and 15 teams also submitted a task description paper.

In future work, we plan to increase the data size and to add more propaganda techniques. We further plan to cover several different languages.

%% file: sections/supplemental_material.tex
\section{Data Collection and Annotation}
\label{sec:appendix_annotation_instructions}

\subsection{Data Collection}
To collect the data for the dataset, we used Facebook, as it has many public groups with a large number of users, who intentionally or unintentionally share a large number of memes. 
We used our own private Facebook accounts to crawl the public posts from users and groups. To make sure the resulting feed had a sufficient number of memes, we initially selected some public groups focusing on topics such as politics, vaccines, COVID-19, and gender equality. Then, using the links between groups, we expanded our initial group pool to a total of 26 public groups.
We went through each group, and we collected memes from old posts, dating up to three months before the newest post in the group. 
Out of the 26 groups, 23 were about politics, US and Canadian: left, right, centered, anti-government, and gun control. The other 3 groups were on general topics such as health, COVID-19, pro-vaccines, anti-vaccines, and gender equality.
Even though the number of political groups was larger (i.e., 23), the other 3 general groups had a higher number of users and a substantial amount of memes.

\subsection{Annotation Process}

We annotated the memes using the 22 persuasion techniques from Section~\ref{sec:annotation} in a multi-label setup. Our annotation focused (\emph{i})~on the text only, using 20 techniques, and (\emph{ii})~on the entire meme (text + image), using all 22 techniques.

We could not annotate the visual modality as an independent task because memes have the text as part of the image. 
Moreover, in many cases, the message in the meme requires both modalities. For example, in Figure~\ref{fig:hitlerum}, the image by itself does not contain any persuasion technique, but together with the text, we can see \emph{Smears} and \emph{Reductio at Hitlerum}.

The annotation team included six members, both female and male, all fluent in English, with qualifications ranging from undergrad to MSc and PhD degrees, including experienced NLP researchers, and covering multiple nationalities. This helped to ensure the quality of the annotation, and our focus was really on having very high-quality annotation. No incentives were given to the annotators. 

\begin{figure*}[!tbh]
\centering
\includegraphics[width=1.0\textwidth]{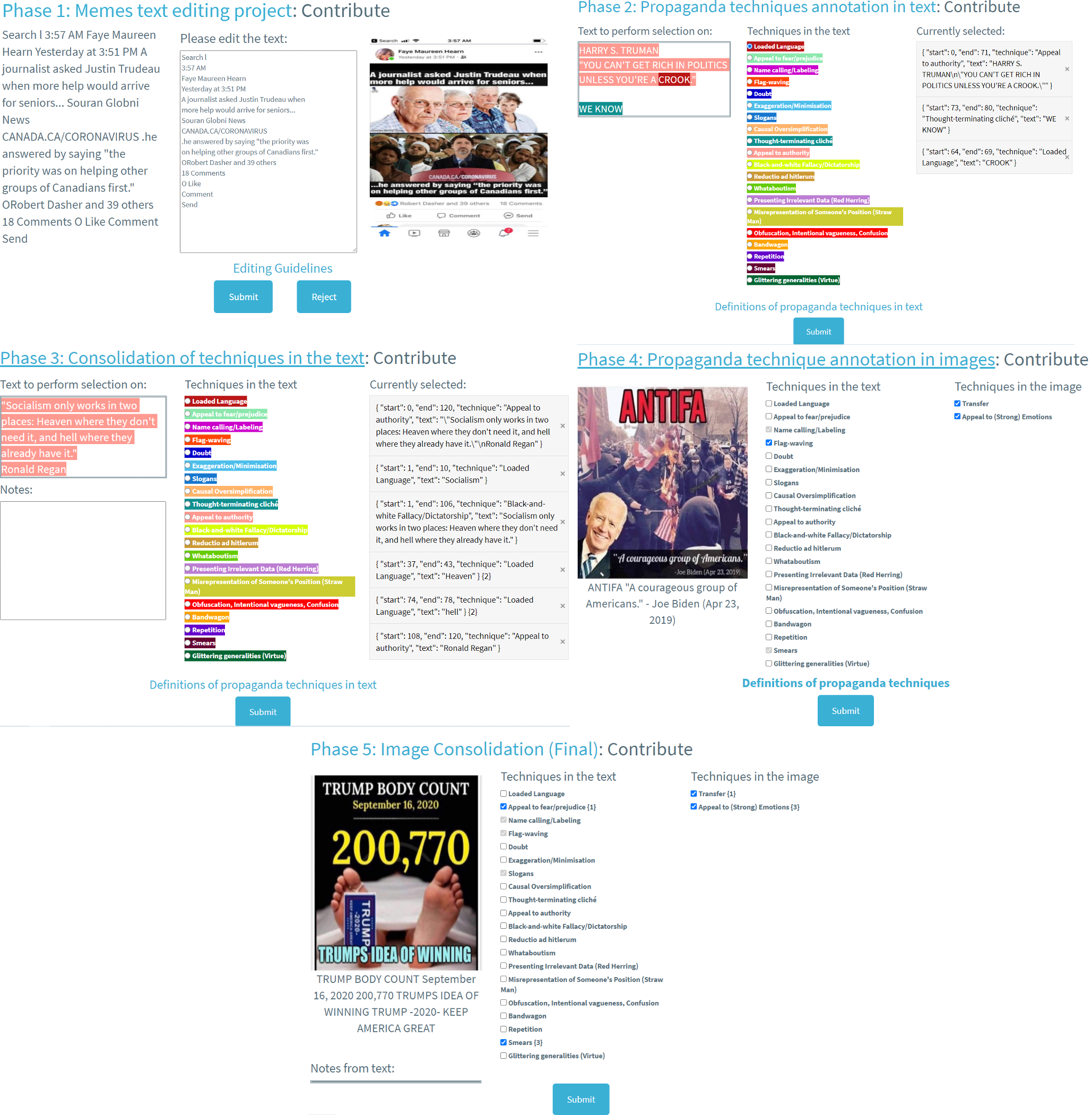}
\caption{Examples of the annotation interface for different phases.}
\label{fig:pyb_example}
\end{figure*}

We used PyBossa\footnote{\url{https://pybossa.com}} as an annotation platform, as it provides the functionality to create a custom annotation interface that we found to be a good fit for our needs in each phase of the annotation process. Figure~\ref{fig:pyb_example} shows examples of the annotation interface for the five different phases of annotation, which we describe in detail below.

\paragraph{Phase 1: Filtering and Text Editing} The first phase of the annotation process is about selecting the memes for our task, followed by extracting and editing the textual contents of each meme. After we collected the memes, we observed that we needed to remove some of them as they did not fit our definition: ``\textit{photograph style image with a short text on top of it}.'' Thus, we asked the annotators to exclude images with the characteristics listed below. During this phase, we filtered out a total of 111 memes.

\begin{figure}[!tbh]
\centering
    \begin{subfigure}[b]{0.48\textwidth}    
        \includegraphics[width=\textwidth]{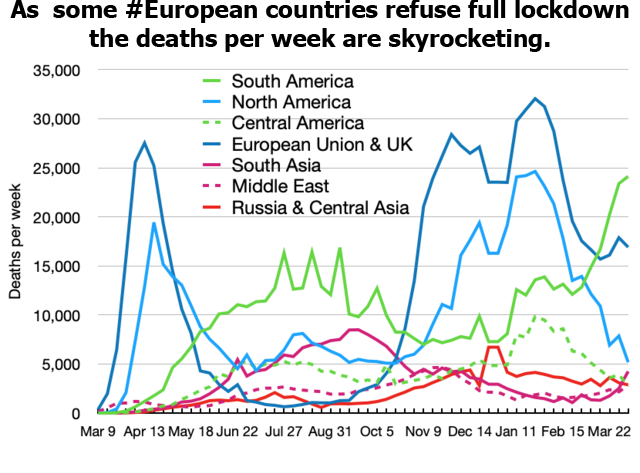}
	    \caption{Example of a meme with a \textbf{graph} \msrc 
	    \imgsrc{https://commons.wikimedia.org/wiki/File:Regional_Covid-19_deaths.png}{}; \ccfrth{}
	    }
    	\label{fig:table_rejection}
    \end{subfigure}    
    \hfill            
    \begin{subfigure}[b]{0.48\textwidth}
    	\includegraphics[width=\textwidth]{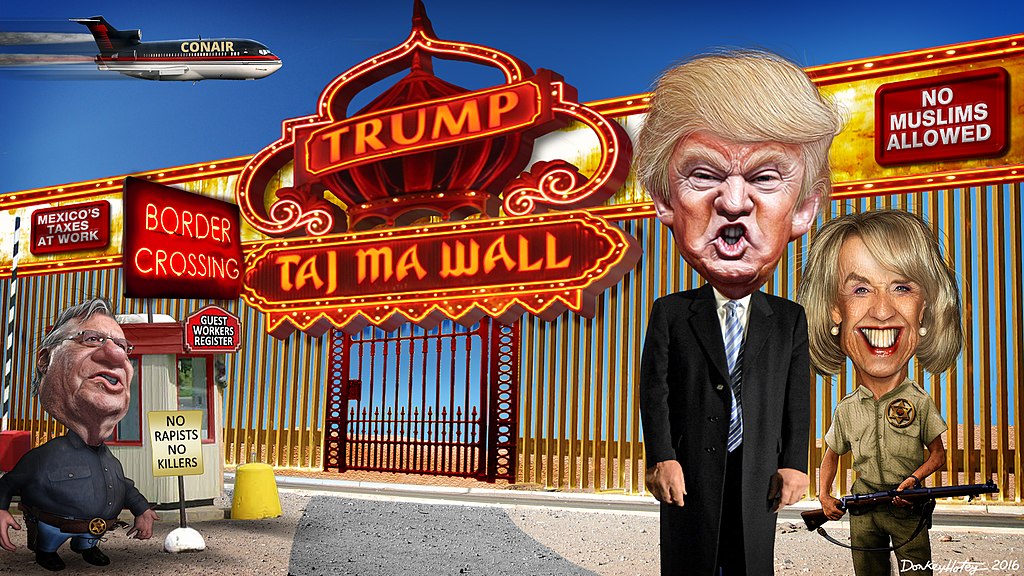}
	    \caption{Example of a \textbf{cartoon} meme;\msrc \imgsrc{https://commons.wikimedia.org/wiki/File:Donald_Trump\%27s_Taj_Ma_WALL_(25844929782).jpg}{}; \ccsnd{}. 
	    }
    	\label{fig:cartoon_rejection}
    \end{subfigure}
    \hfill
    \begin{subfigure}[b]{0.48\textwidth}
        \includegraphics[width=\textwidth]{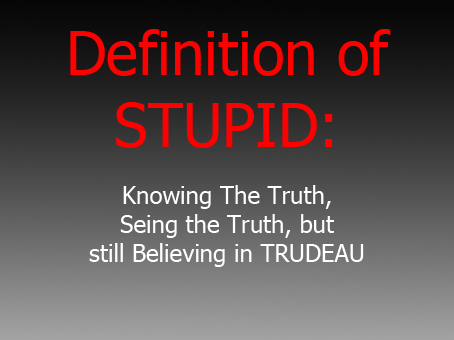}
    	\caption{Example of a meme with \textbf{only text modality}; \public{}. 
    	}
	    \label{fig:unimodal_rejection}
    \end{subfigure}    
    \caption{Examples of memes we filtered out.}
    \label{fig:memes_example_for_filtering}
\end{figure}

\begin{itemize}
    \item Images with diagrams/graphs/tables (see Figure \ref{fig:table_rejection}).
    \item Cartoons. (see Figure \ref{fig:cartoon_rejection})
    \item Memes for which no multi-modal analysis is possible: e.g., only text, only image, etc. (see Figure \ref{fig:unimodal_rejection})
\end{itemize}

Next, we used the Google Vision API\footnote{\url{http://cloud.google.com/vision}} to extract the text from the memes. As the resulting text sometimes contains errors, manual checking was needed to correct it. Thus, we defined several text editing rules, and we asked the annotators to apply them on the memes that passed the filtering rules above.

\begin{enumerate}
\item When the meme is a screenshot of a social network account, e.g., WhatsApp, the user name and login can be removed as well as all ``Like'', ``Comment', ``Share''.
\item Remove the text related to logos that are not part of the main text.
\item Remove all text related to figures and tables.
\item Remove all text that is partially hidden by an image, so that the sentence is almost impossible to read.
\item Remove all text that is not from the meme, but on banners carried on by demonstrators, street advertisements, etc.
\item Remove the author of the meme if it is signed.
\item If the text is in columns, first put all text from the first column, then all text from the next column, etc.
\item Rearrange the text, so that there is one sentence per line, whenever possible.
\item If there are separate blocks of text in different locations of the image, separate them by a blank line. However, if it is evident that the text blocks are part of a single sentence, keep them together.
\end{enumerate}

\paragraph{Phase 2: Text Annotation} The annotations for phase 2 are targeted at Subtasks 1 and 2. Given the list of propaganda techniques for text only annotation, as discussed in Section~\ref{ssec:appendix_definitions_prop_tech} (i.e.,~techniques 1-20), and the textual content of the target meme, the annotators were asked to identify which techniques appear in the text, and also to annotate the span of each instance of a technique use. 
In this phase, there were three annotators per example.

\paragraph{Phase 3: Text Consolidation} Phase 3 is the consolidation step for the annotations from phase 2. The three annotators met with the rest of the team, who acted as consolidators, and discussed each annotation, so that a consensus could be reached. 

We made sure to consider different interpretations and to anotate techniques corresponding to the most likely one. While this phase was devoted to checking the annotations from phase 2, when a novel instance of a technique was found, it could be added; conversely, an instance of a technique with perfect agreement from phase 2 could also be dropped. Phase 3 was essential for ensuring quality, and it served as an additional training opportunity for the entire team, which was very useful. 

\paragraph{Phase 4: Multimodal Annotation}
In this phase, the goal is to identify which of the 22 techniques, discussed in Section~\ref{ssec:appendix_definitions_prop_tech}, appear in the meme: in the text and in the visual content. Note that some of the techniques occurring in the text might be identified only in this phase because the image provides the necessary context. 
Thus, we presented the meme with the consolidated propaganda labels from phase 3.
We intentionally provided the consolidated text labels to the annotators in order to ensure that they focus their attention on identifying propaganda techniques that require both modalities rather than repeating what was already labeled in the earlier phases.
In this phase, there were three annotators per example.

\paragraph{Phase 5: Multimodal Consolidation.} In phase 5, we consolidated the annotations from phase 4 in a discussion of the entire team of six annotators (just as we did for phase 3).

\subsection{Annotation Agreement}
We assessed the quality for the individual annotators from phases 2 and 4 (i.e., when combining the annotations for the meme's text and for the entire meme) to the final consolidated labels at phase 5. Since our annotation is multi-label, we computed Krippendorff's $\alpha$ \cite{artstein2008inter}. The results are shown in Table~\ref{tab:annotation_agr}, and the numbers indicate moderate to substantial agreement \cite{landis1977measurement}.
 
\begin{table}[tbh]
\centering
\scalebox{0.94}{
\begin{tabular}{lc}
\toprule
\bf Agreement Pair & \bf Krippendorff's $\alpha$ \\
\midrule
Annotator 1 vs. Consolidated & 0.83 \\
Annotator 2 vs. Consolidated & 0.91 \\
Annotator 3 vs. Consolidated & 0.56 \\
\midrule
\textbf{Average} & \textbf{0.77} \\
\bottomrule
\end{tabular}
}
\caption{Inter-annotator agreement in terms of Krippendorff's $\alpha$ between each of the annotators and the consolidated annotation.}
\label{tab:annotation_agr}
\end{table}

\subsection{Propaganda Techniques: Definitions}
\label{ssec:appendix_definitions_prop_tech}

Below, we present the definitions of our 22 propaganda techniques, together with examples: both textual, and memes. Note that, for copyright reasons, we show our own recreated versions of actual memes from our dataset, where, for each meme, we indicate the image(s) we used and the corresponding license terms (as hyperlinks in the image caption).

\textbf{~\\1. Loaded Language:} 
Using specific words and phrases with strong emotional implications (i.e., either positive or negative) to influence an audience. 

\begin{figure}[!tbh]
    \centering
    \includegraphics[width=0.97\columnwidth]{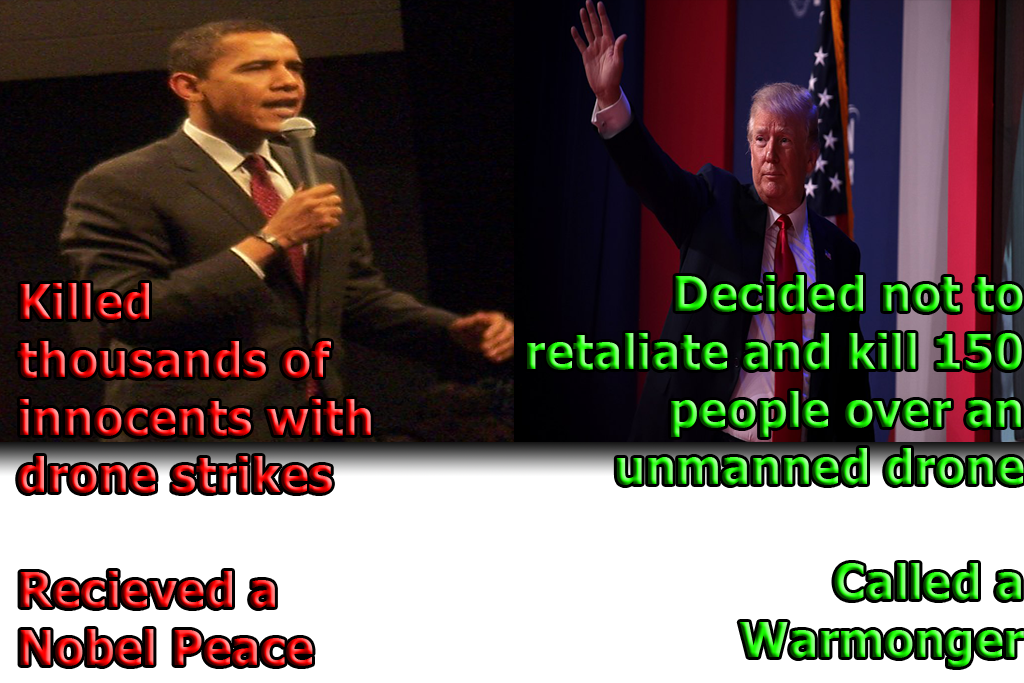}
    \caption{Example for \textbf{Loaded Language};\msrc
    \imgsrc{https://commons.wikimedia.org/wiki/File:OBAMA_165.JPG}{1},
    \imgsrc{https://commons.wikimedia.org/wiki/File:Donald_Trump_(40525851791).jpg}{2};
    \public{1},
    \ccsnd{2}
    }
    \label{fig:loaded_exmpl}
\end{figure}

An example meme is shown in Figure \ref{fig:loaded_exmpl}, which contains four instances of this persuasion technique in its text: \emph{killed thousands of innocents}, \emph{retaliate}, \emph{kill}, and \emph{warmonger}.

\textbf{~\\2. Name Calling or Labeling:}
Labeling the object of the propaganda as either something the target audience fears, hates, finds undesirable, or loves, praises. 

\begin{figure}[!h]
    \centering
    \includegraphics[width=0.97\columnwidth]{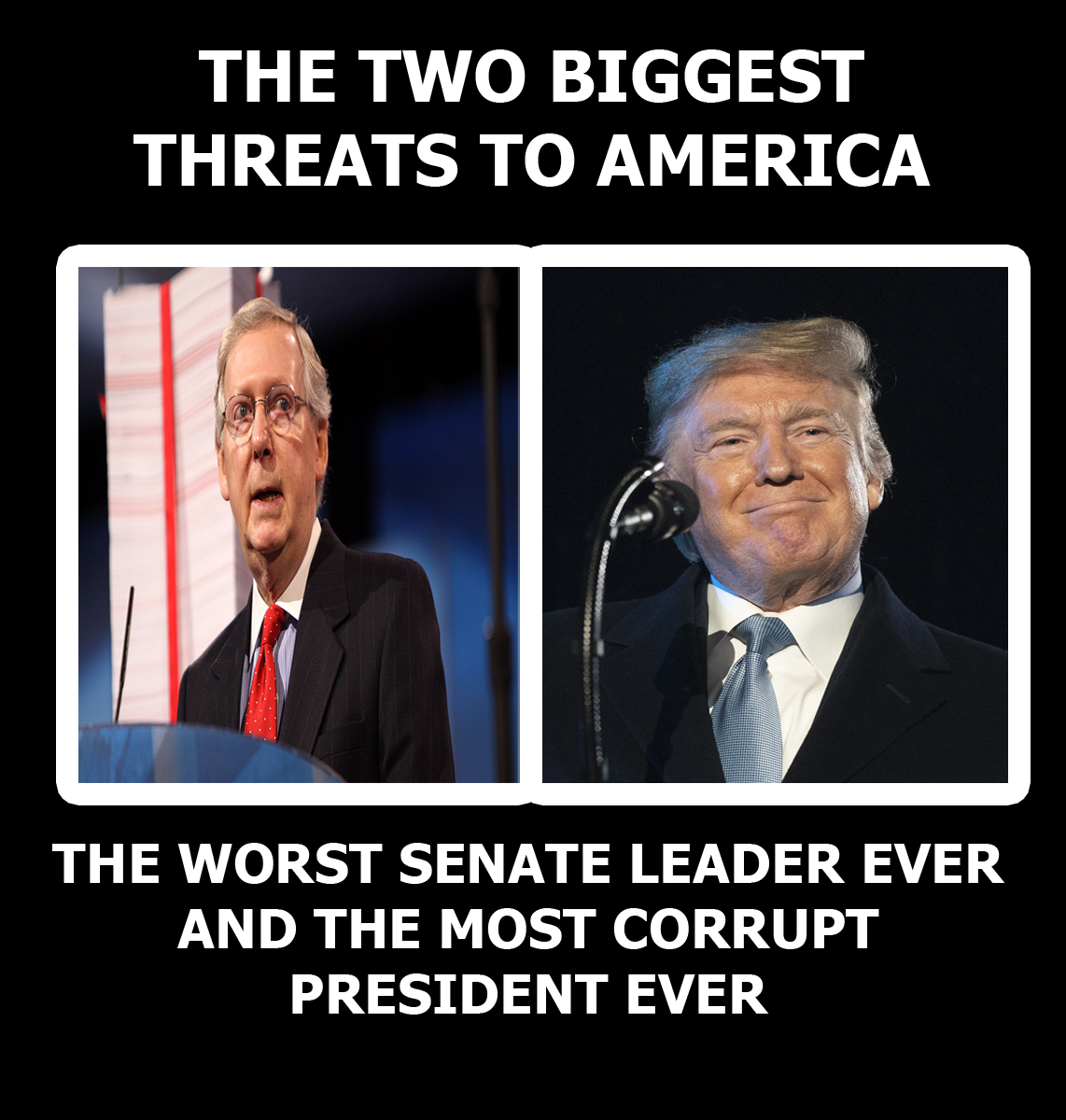}
    \caption{Example for \textbf{Name Calling};\msrc 
    \imgsrc{https://commons.wikimedia.org/wiki/File:Mitch_McConnell_(8567871168).jpg}{1},
    \imgsrc{https://commons.wikimedia.org/wiki/File:Donald_Trump_at_2019_National_Christmas_Tree_Lighting_Ceremony_(49178332061)_(cropped).jpg}{2};
    \public{1},
    \ccsnd{2}
    }
    \label{fig:calling_exmpl}
\end{figure}

Figure~\ref{fig:calling_exmpl} shows three instances of this technique: \emph{the two biggest threats to America}, \emph{the worst senate leader ever}, and \emph{the most corrupt President ever}. Figure~\ref{fig:loaded_exmpl} also contains an instance: \emph{warmonger}.

\textbf{~\\3. Doubt:}
Questioning the credibility of someone or something. 

\begin{figure}[h!]
    \centering
    \includegraphics[width=0.9\columnwidth]{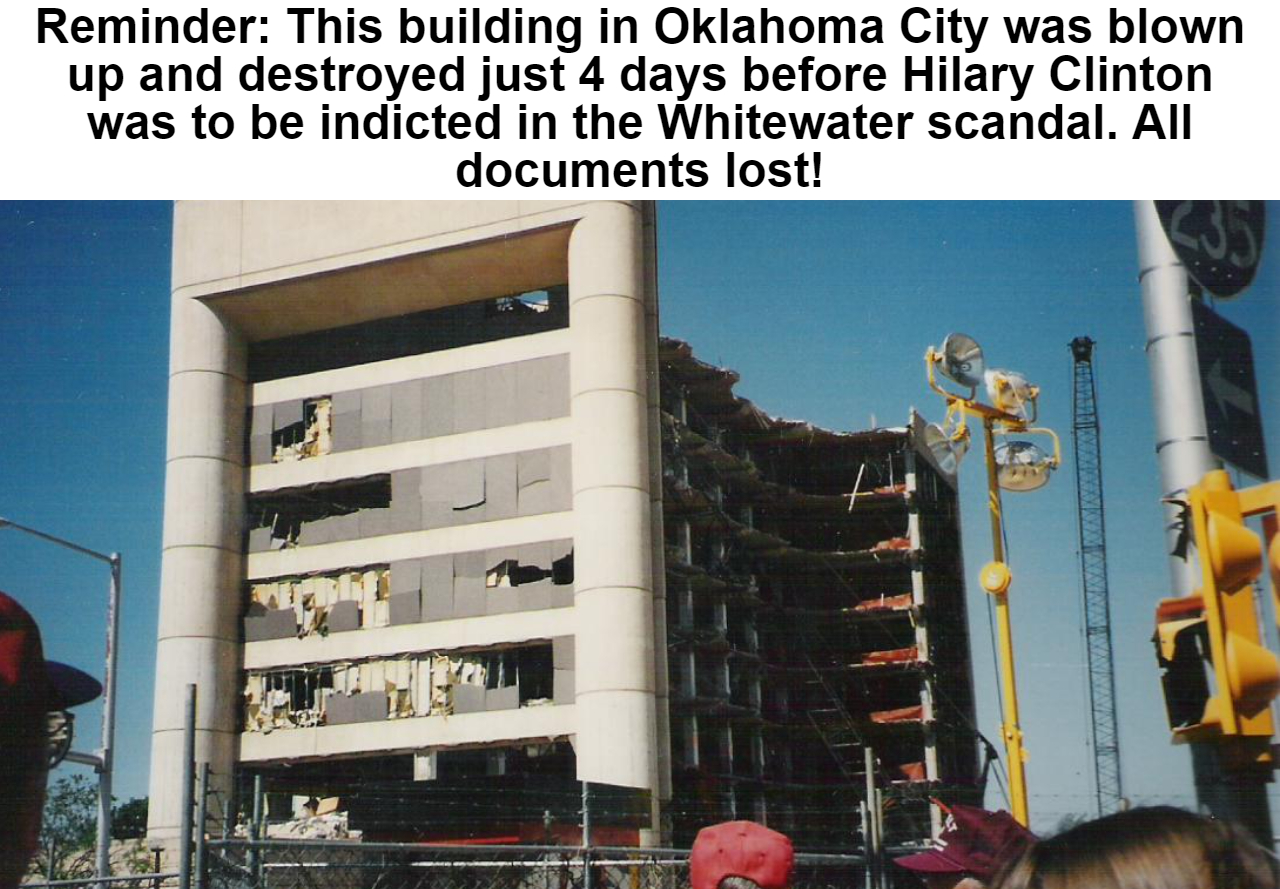}
    \caption{Example for \textbf{Doubt};\msrc 
    \imgsrc{https://commons.wikimedia.org/wiki/File:Murrah_Building_Before_Demolition.JPG}{};
    \public{}
    }
    \label{fig:doubt_exmple}
\end{figure}

An example is shown in Figure \ref{fig:doubt_exmple}, where the entire text in the meme represents a span for this technique, while the image is just for illustration.

\textbf{~\\4. Exaggeration or Minimisation:}
Representing something in an excessive manner, making it larger, better, worse (e.g., \emph{the best of the best}); or making it seem less important or smaller than it really is (e.g., saying that an insult was just a joke). 

\begin{figure}[h!]
    \centering
    \includegraphics[width=0.80\columnwidth]{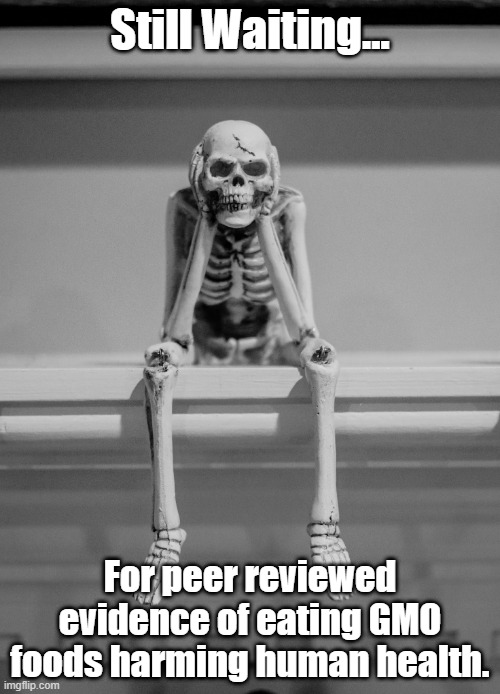}
    \caption{Example for \textbf{Exaggeration};\msrc 
    \imgsrc{https://images.unsplash.com/photo-1540264709335-e6c6be3e1d71?ixid=MXwxMjA3fDB8MHxwaG90by1wYWdlfHx8fGVufDB8fHw\%3D&ixlib=rb-1.2.1&auto=format&fit=crop&w=750&q=80}{};
    \unplash{}
    }
    \label{fig:exagg_exmpl}
\end{figure}

An example is shown in Figure~\ref{fig:exagg_exmpl}, where the entire meme conveys an exaggeration. Moreover, all three \emph{Name Calling} instances in Figure~\ref{fig:calling_exmpl} are also examples of \emph{Exaggeration}.

\textbf{~\\5. Appeal to Fear/Prejudice:} 
Seeking to build support for an idea by instilling anxiety and/or panic in the population towards an alternative. In some cases, the support is built based on preconceived judgments. 

\begin{figure}[h!]
    \centering
    \includegraphics[width=1.0\columnwidth]{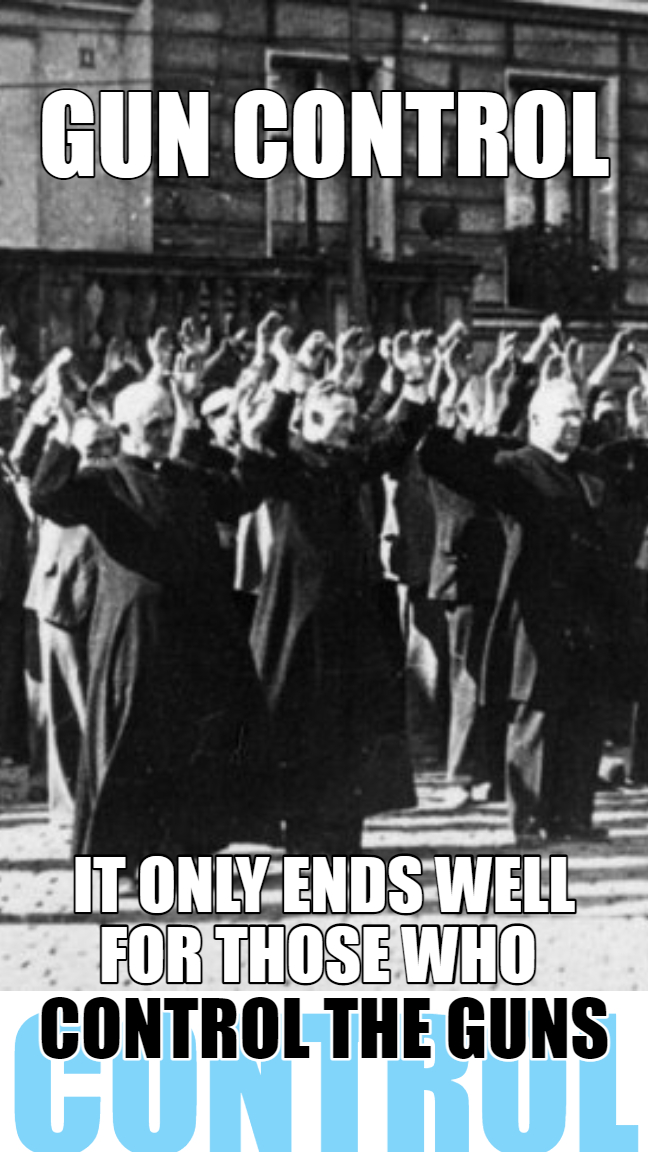}
    \caption{Example for \textbf{Appeal to Fear};\msrc
    \imgsrc{https://commons.wikimedia.org/wiki/File:Bydgoszcz_1939_Polish_priests_and_civilians_at_the_Old_Market.jpg}{};
    \public{}
    }
    \label{fig:fear_exmpl}
\end{figure}

An example is shown in Figure~\ref{fig:fear_exmpl}, where both the text and the image instill fear. 

\newpage
\textbf{~\\6. Slogans:}
A brief and striking phrase that may include labeling and stereotyping. Slogans tend to act as emotional appeals. 

\begin{figure}[!tbh]
    \centering
    \includegraphics[width=\columnwidth]{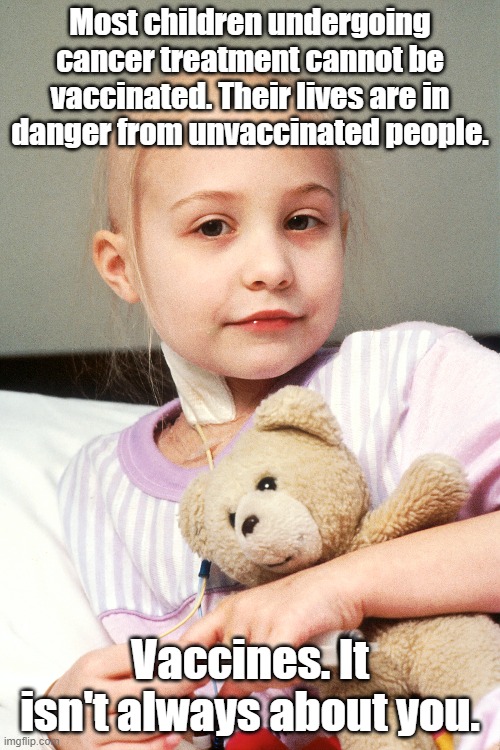}
    \caption{Example for \textbf{Slogan};\msrc 
    \imgsrc{https://images.unsplash.com/photo-1578496781197-b85385c7f0b3?ixlib=rb-1.2.1&ixid=MXwxMjA3fDB8MHxwaG90by1wYWdlfHx8fGVufDB8fHw\%3D&auto=format&fit=crop&w=634&q=80}{};
    \unplash{}
    }
    \label{fig:slogan_exmpl}
\end{figure}

An example is shown in Figure~\ref{fig:slogan_exmpl}, which contains a slogan in its textual content: ``\emph{Vaccines. It isn't always about you.}''

\newpage
\textbf{~\\7. Whataboutism:}
A technique that attempts to discredit an opponent's position by charging them with hypocrisy without directly disproving their argument. 

\begin{figure}[!tbh]
    \centering
    \includegraphics[width=\columnwidth]{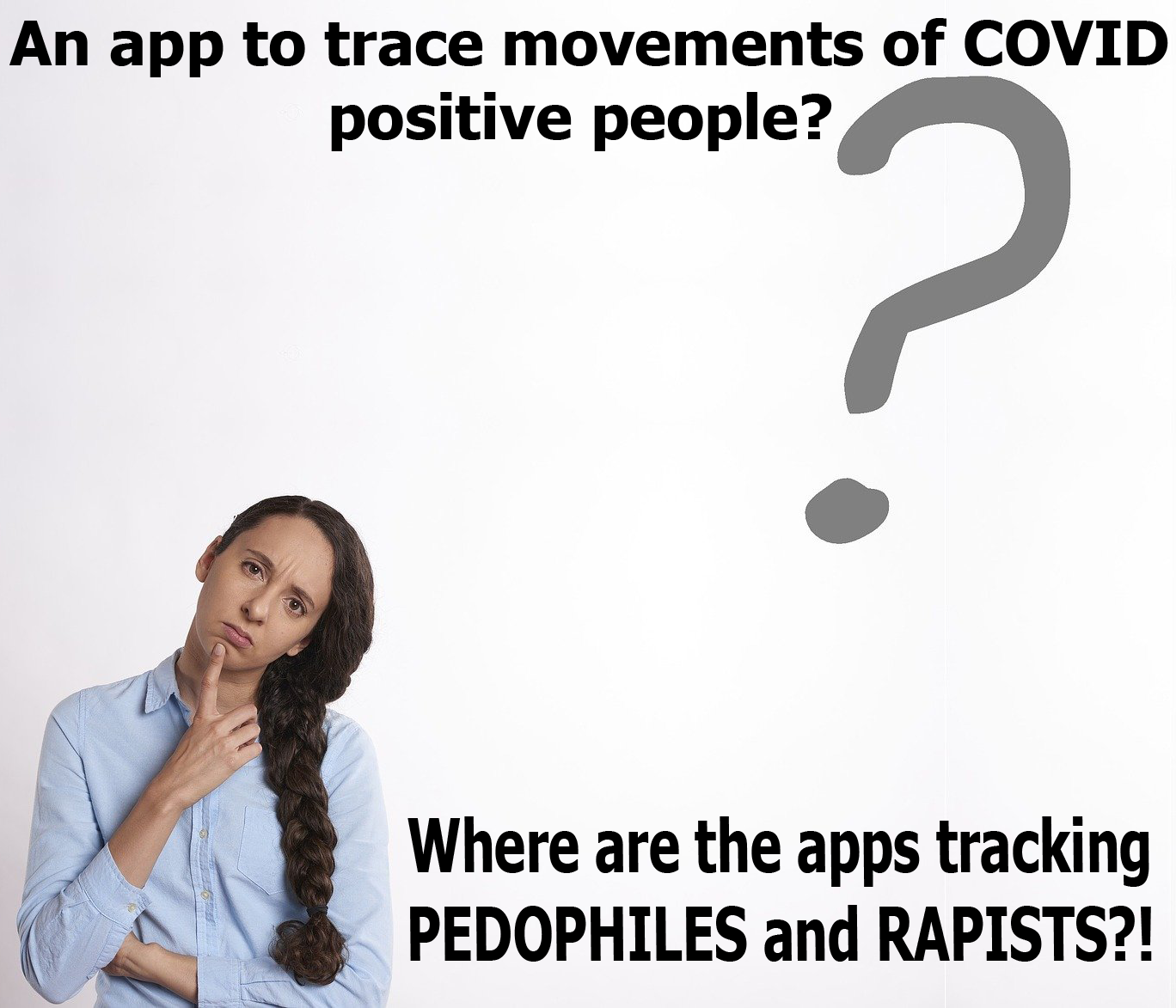}
    \caption{Example for \textbf{Whataboutism};\msrc 
    \imgsrc{https://pixabay.com/photos/question-doubt-problem-mark-3385451/}{};
    \pixelbay{}
    }
    \label{fig:whata_exmpl}
\end{figure}

An example meme is shown in Figure~\ref{fig:whata_exmpl}, where the entire text represents a span for this technique, while the image is just for illustration.

\newpage
\textbf{~\\8. Flag-Waving:}
Playing on strong national feeling (or to any group such as race, gender, political preference) to justify or promote an action or idea. 

\begin{figure}[h!]
    \centering
    \includegraphics[width=\columnwidth]{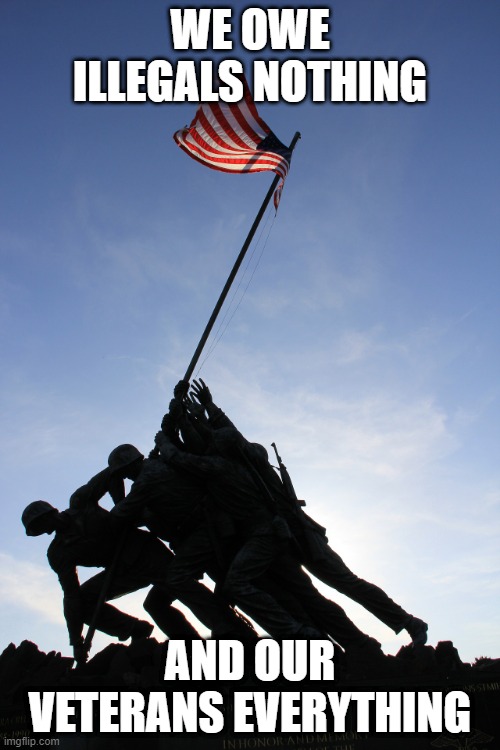}
    \caption{Example for  \textbf{Flag-Waving};\msrc 
    \imgsrc{https://cdn.pixabay.com/photo/2014/06/06/21/06/statue-363846_1280.jpg}{};
    \pixelbay{}
    }
    \label{fig:flag_exmpl}
\end{figure}

An example is shown in Figure~\ref{fig:flag_exmpl}, with the technique expressed in the text and the image.

\newpage
\textbf{~\\9. Misrepresentation of Someone's Position (Straw Man):}
An opponent's proposition is substituted with a similar one, which is then refuted in place of the original proposition. 

\begin{figure}[!tbh]
    \centering
    \includegraphics[width=\columnwidth]{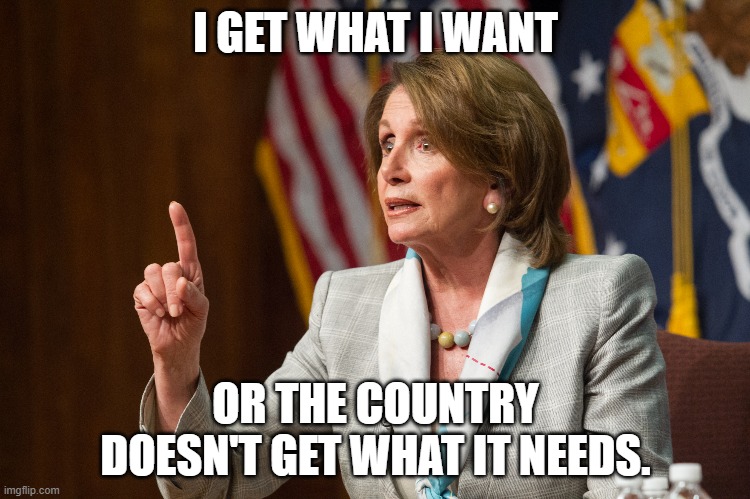}
    \caption{Example for \textbf{Misrepresentation of Someone's Position (Straw Man)};\msrc 
    \imgsrc{https://commons.wikimedia.org/wiki/File:L-15-04-14-A.653_(16941906367).jpg}{};
    \public{}
    }
    \label{fig:straw_exmpl}
\end{figure}

An example meme is shown in Figure~\ref{fig:straw_exmpl}, which contains an instance of this technique in its text: here, the entire text in the meme represents a span for this technique, while the image is irrelevant for that technique (however, it is relevant for other techniques such as \emph{Smears}).

\newpage
\textbf{~\\10. Causal Oversimplification:}
Assuming a single cause or reason when there are actually multiple causes for an issue. It includes transferring blame to one person or group of people without investigating the complexities of the issue. 
 
\begin{figure}[!tbh]
    \centering
    \includegraphics[width=\columnwidth]{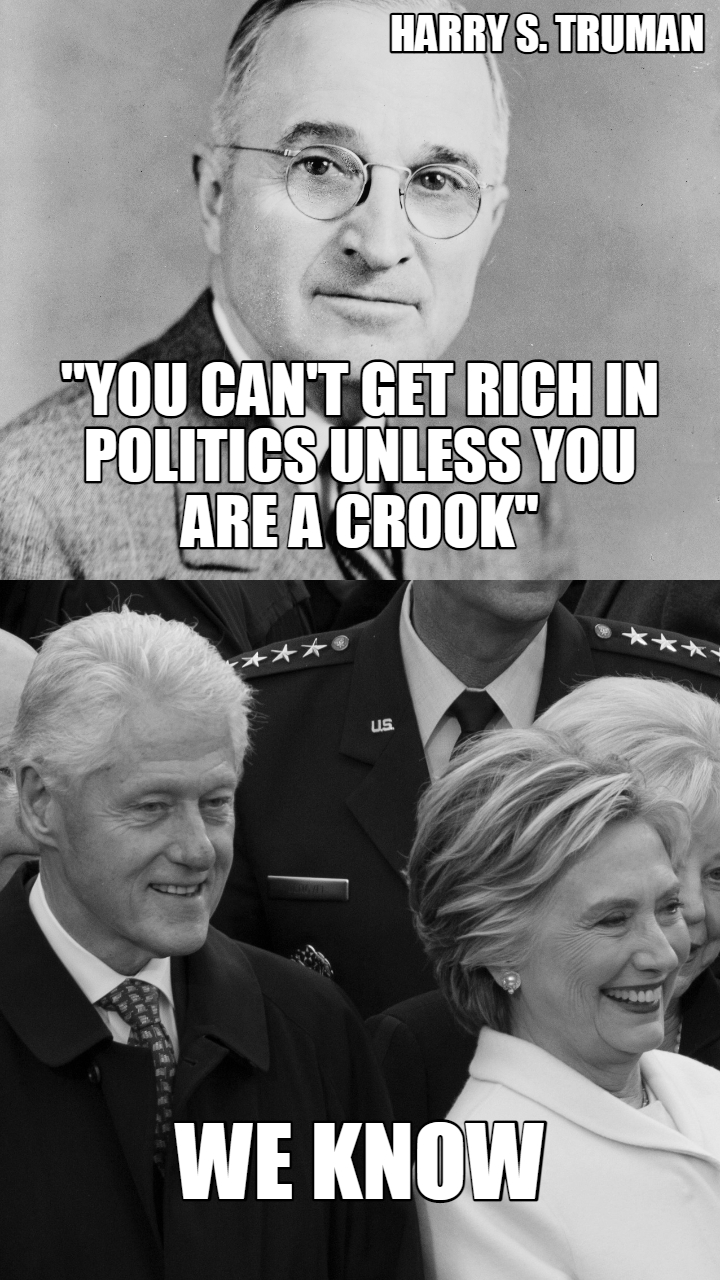}
    \caption{Example for \textbf{Causal Oversimplification};\msrc 
     \imgsrc{https://images.unsplash.com/photo-1580128636867-7224f71904fd?ixid=MXwxMjA3fDB8MHxwaG90by1wYWdlfHx8fGVufDB8fHw\%3D&ixlib=rb-1.2.1&auto=format&fit=crop&w=696&q=80}{1},
    \imgsrc{https://commons.wikimedia.org/wiki/File:Bill_and_Hillary_Clinton_at_58th_Inauguration_01-20-17_(cropped).jpg}{2};
    \unplash{1},
    \public{2}
    }
    \label{fig:oversimpl_exmpl}
\end{figure}

An example meme is shown in Figure~\ref{fig:oversimpl_exmpl}, which contains an instance of this technique in its text: ``\emph{You can't get rich in politics unless you are a crook.}'' This statement says that if somebody got rich in politics, the only reason for this happening should be that this person is a crook, while in reality there are typically multiple causes.
The image is irrelevant for that technique (however, it is relevant for other techniques such as \emph{Smears}).

\newpage
\textbf{~\\11. Appeal to Authority:}
Stating that a claim is true simply because a valid authority or expert on the issue said it was true, without any other supporting evidence offered. We consider the special case in which the reference is not an authority or an expert in this technique, although it is referred to as \textit{Testimonial} in literature. 

\begin{figure}[!tbh]
    \centering
    \includegraphics[width=\columnwidth]{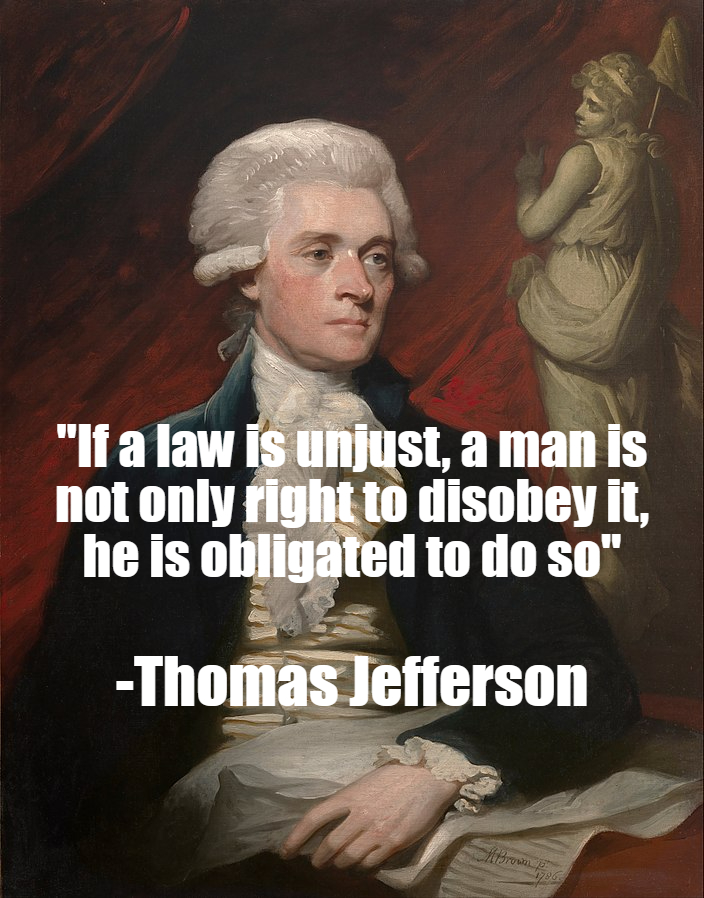}
    \caption{Example for \textbf{Appeal to Authority};\msrc
    \imgsrc{https://commons.wikimedia.org/wiki/File:Mather_Brown_-_Thomas_Jefferson_-_Google_Art_Project.jpg}{};
    \public{}
    }
    \label{fig:authority_exmpl}
\end{figure}

An example meme is shown in Figure \ref{fig:authority_exmpl}, which contains a quote by the 3rd President of the United States.

\newpage
\textbf{~\\12. Thought-Terminating Clich\'e:}
Words or phrases that discourage critical thought and meaningful discussion about a given topic. They are typically short, generic sentences that offer seemingly simple answers to complex questions or that distract attention away from other lines of thought. 

\begin{figure}[h!]
    \centering
    \includegraphics[width=\columnwidth]{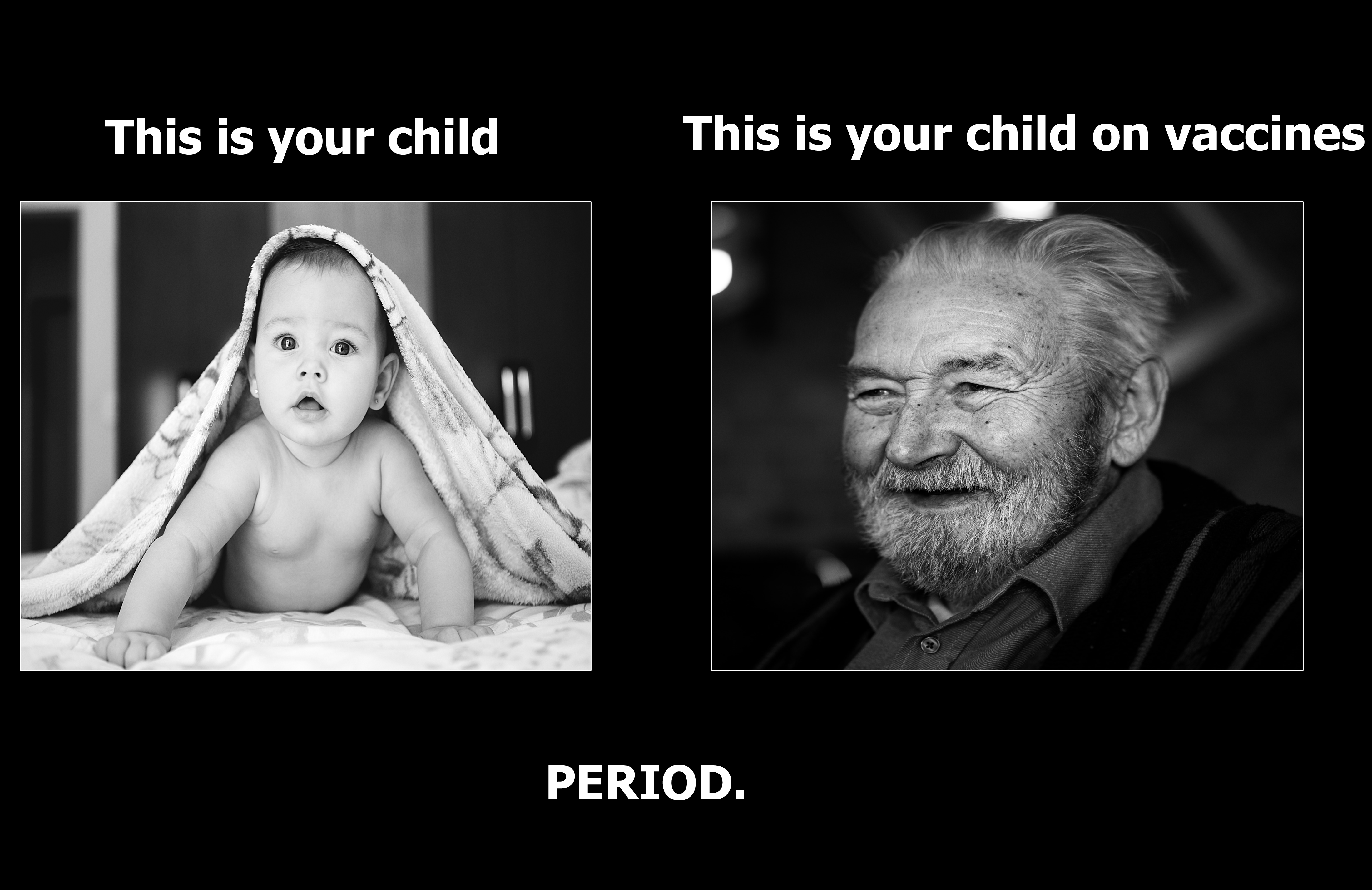}
    \caption{Example for  \textbf{Thought-Terminating Clich\'{e}};\msrc 
    \imgsrc{https://unsplash.com/photos/CgWTqYxHEkg}{1},
    \imgsrc{https://unsplash.com/photos/BBjW2qnIixc}{2};
    \unplash{1},
    \unplash{2}
    }
    \label{fig:cliche_exmpl}
\end{figure}

Figure \ref{fig:cliche_exmpl} shows a meme with an instance of this technique in its text: ``\emph{PERIOD}.''

\textbf{~\\13. Black-and-White Fallacy:}
Presenting two alternative options as the only possibilities, when in fact more possibilities exist. We also include dictatorship, where one tells the audience exactly what actions to take, eliminating any other choices. 

\begin{figure}[h!]
    \centering
    \includegraphics[width=0.88\columnwidth]{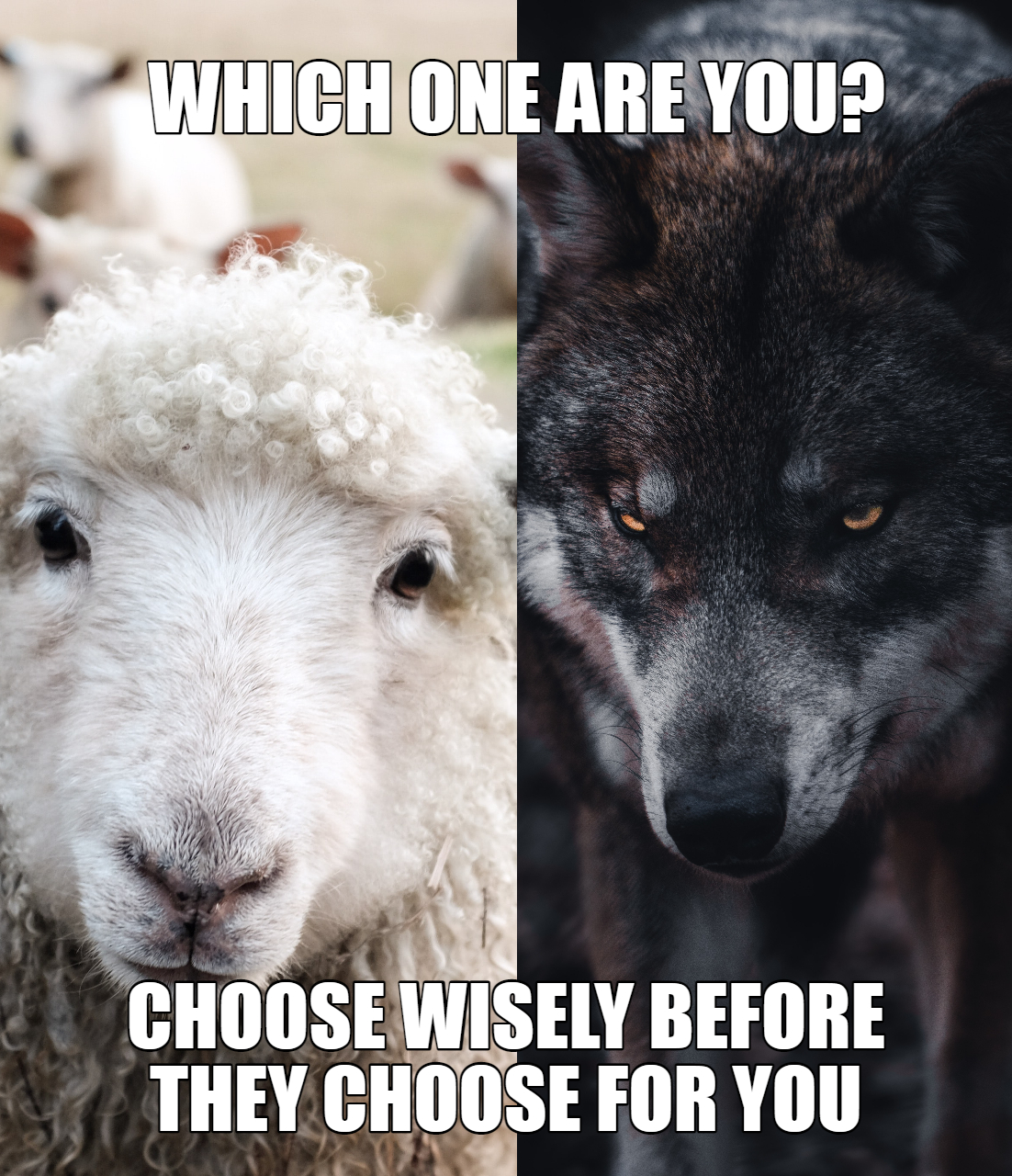}
    \caption{Example for \textbf{Black-and-White Fallacy};\msrc
    \imgsrc{https://images.unsplash.com/photo-1484557985045-edf25e08da73?ixlib=rb-1.2.1&ixid=MXwxMjA3fDB8MHxwaG90by1wYWdlfHx8fGVufDB8fHw\%3D&auto=format&fit=crop&w=667&q=80}{1},
    \imgsrc{https://images.unsplash.com/photo-1590420485404-f86d22b8abf8?ixid=MXwxMjA3fDB8MHxwaG90by1wYWdlfHx8fGVufDB8fHw\%3D&ixlib=rb-1.2.1&auto=format&fit=crop&w=634&q=80}{2};
    \unplash{1},
    \unplash{2}
    }
    \label{fig:blnwhite_exmpl}
\end{figure}

An example of this technique is shown in Figure~\ref{fig:blnwhite_exmpl}, which offers only two choices.

\newpage
\textbf{~\\14. Reductio ad Hitlerum:}
Persuading an audience to disapprove an action or idea by suggesting that the idea is popular with groups hated or in contempt by the target audience. It can refer to any person or concept with a negative connotation. 

\begin{figure}[!tbh]
    \centering
    \includegraphics[width=\columnwidth]{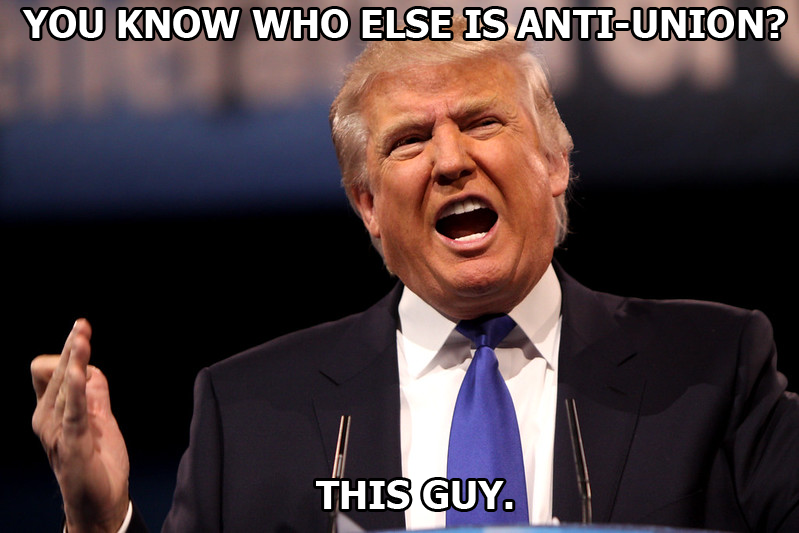}
    \caption{Example for \textbf{Reduction ad Hitlerum};\msrc
    \imgsrc{https://www.flickr.com/photos/22007612@N05/8566717881}{},
    \ccsnd{}
    }
    \label{fig:hitlerum_exmpl}
\end{figure}

Figure \ref{fig:hitlerum_exmpl} shows a meme trying to discredit the idea of being anti-union by saying that so is Donald Trump, who in turn is shown in bad light.

\textbf{~\\15. Repetition:}
Repeating the same message, so that the audience eventually accepts it. 

\begin{figure}[h!]
    \centering
    \includegraphics[width=0.95\columnwidth]{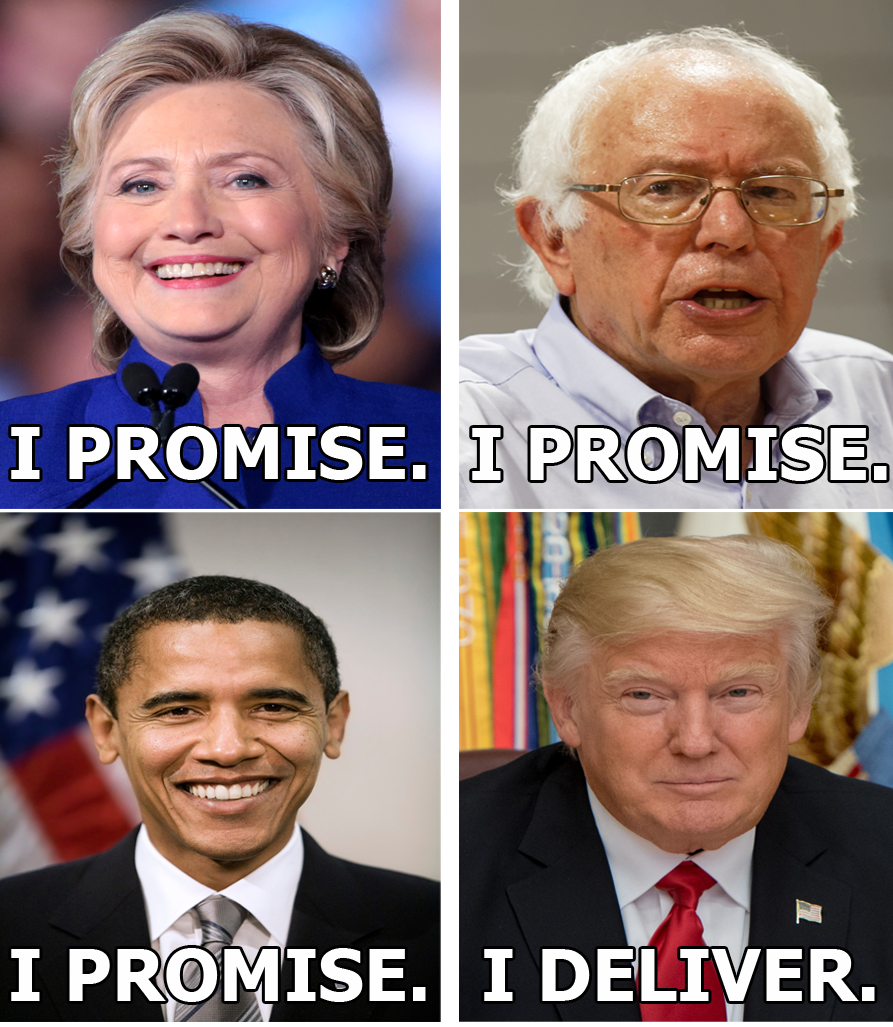}
    \caption{Example for  \textbf{Repetition};\msrc
    \imgsrc{https://commons.wikimedia.org/wiki/File:Donald_Trump_Pentagon_2017.jpg}{1},
    \imgsrc{https://commons.wikimedia.org/wiki/File:Hillary_Clinton_Arizona_2016_.jpg}{2},
    \imgsrc{https://commons.wikimedia.org/wiki/File:US_Senator_of_Vermont_Bernie_Sanders_in_Conway_NH_on_August_24th_2015_by_Michael_Vadon_(20715416790)_(cropped).jpg}{3},
    \imgsrc{https://commons.wikimedia.org/wiki/File:Poster-sized_portrait_of_Barack_Obama_OrigRes.jpg}{4};
     \public{1},
    \ccsnd{2},
    \ccsnd{3},
    \cctrdunprt{4}
    }
    \label{fig:repetition_exmpl}
\end{figure}

An example is shown in Figure~\ref{fig:repetition_exmpl}, where the repetition has a clear rhetorical function.

\textbf{~\\16. Obfuscation, Intentional Vagueness, Confusion:}
Using words that are deliberately unclear, so that the audience may have their own interpretations. 

\begin{figure}[!tbh]
    \centering
    \includegraphics[width=\columnwidth]{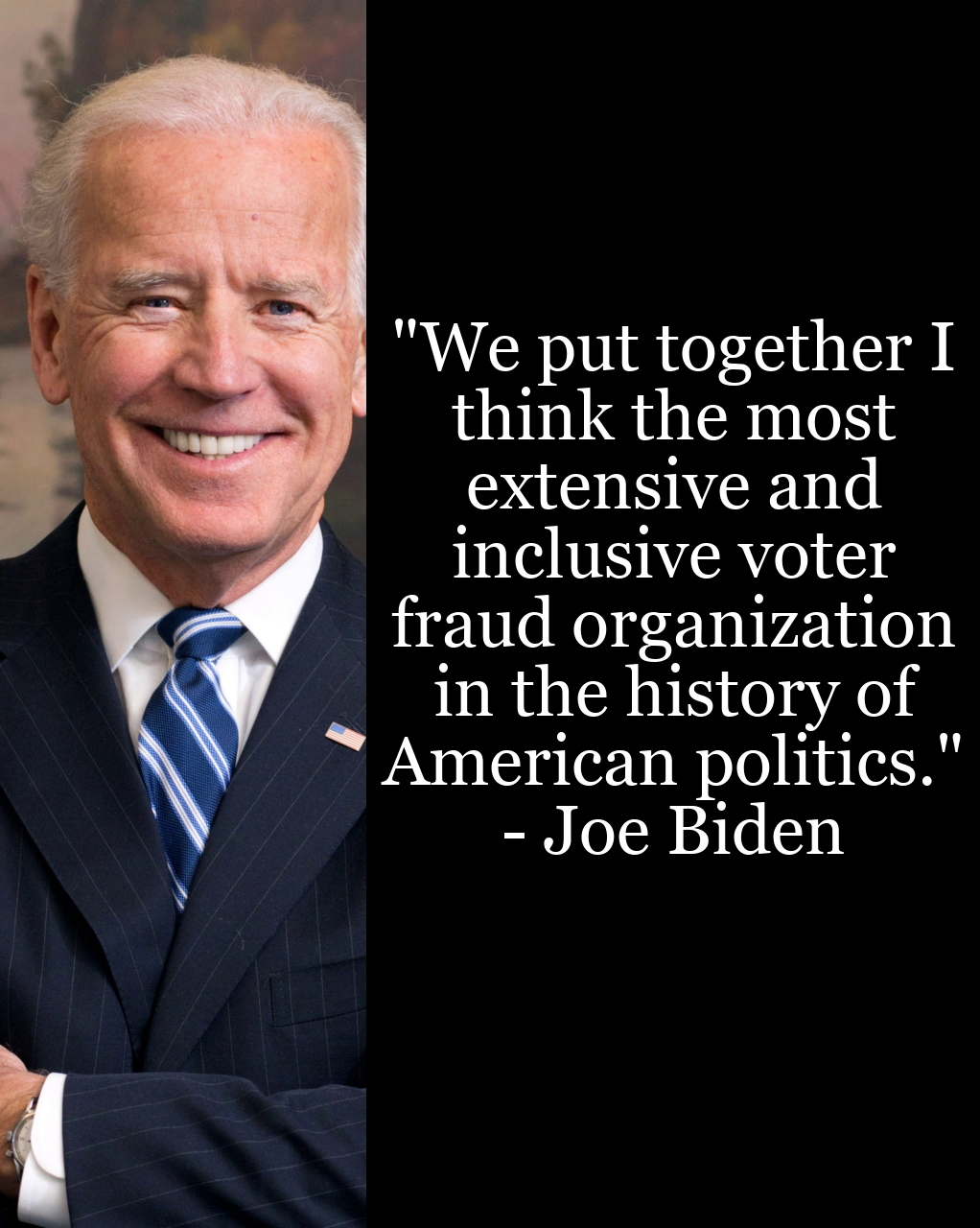}
    \caption{Example for  \textbf{Obfuscation, Intentional vagueness, Confusion};\msrc
    \imgsrc{https://commons.wikimedia.org/wiki/File:Joe_Biden_official_portrait_2013_cropped.jpg}{};
    \public{}
    }
    \label{fig:confusion_exmpl}
\end{figure}

Figure \ref{fig:confusion_exmpl}, shows an example, where the entire quote by Joe Biden is a span of this technique, as it is unclear what exactly is meant here.

\newpage
\textbf{~\\17. Presenting Irrelevant Data (Red Herring):}
Introducing irrelevant material to the issue being discussed, so that everyone's attention is diverted away from the points made. 

\begin{figure}[!tbh]
    \centering
    \includegraphics[width=\columnwidth]{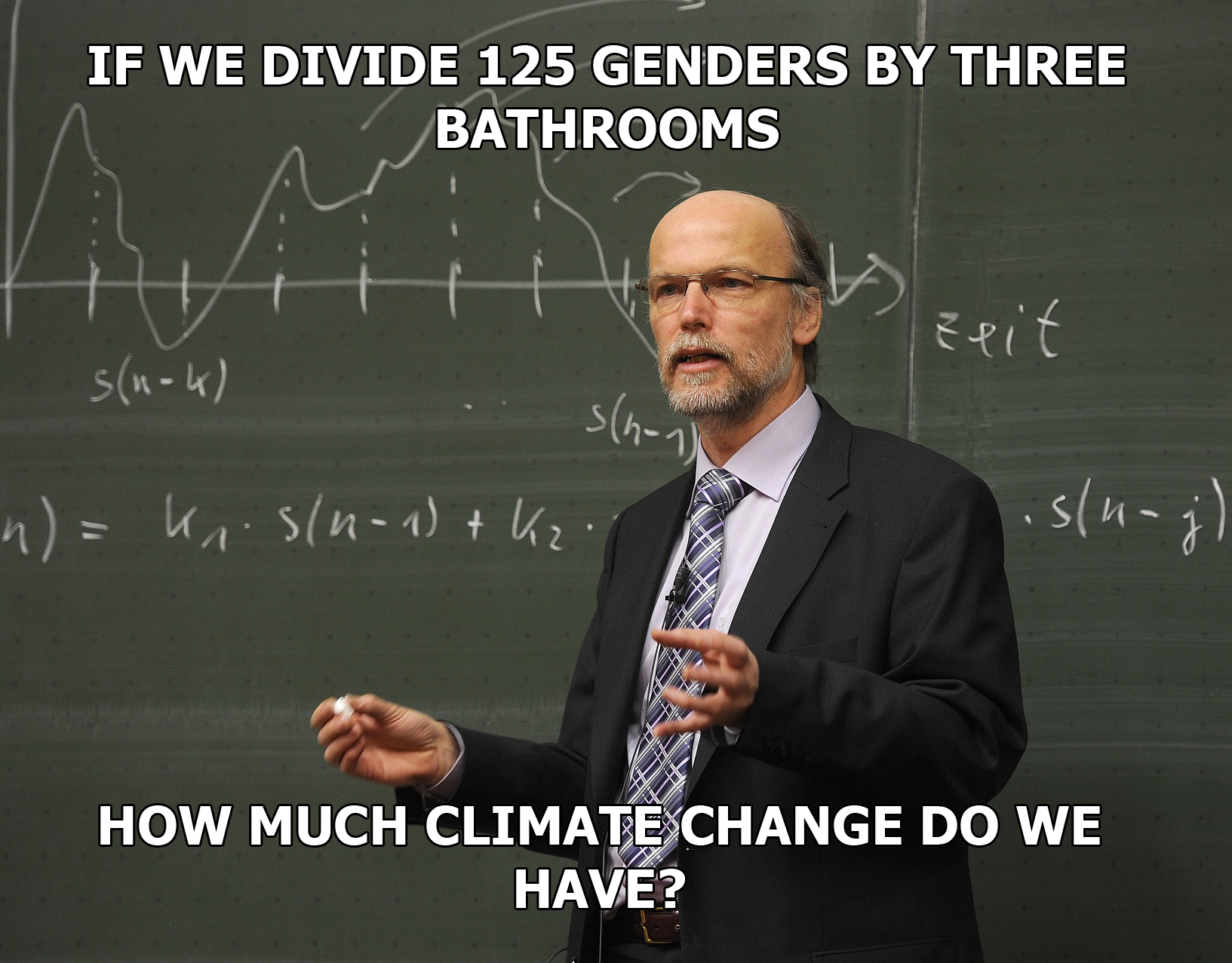}
    \caption{Example for \textbf{Presenting Irrelevant Data (Red Herring)};\msrc  \imgsrc{https://pixabay.com/photos/birger-kollmeier-professor-910261/}{};
    \pixelbay{}
    }
    \label{fig:herring_exmpl}
\end{figure}

An example meme is shown in Figure~\ref{fig:herring_exmpl}, which contains an instance of this technique in its text. We can see that there is no real connection between the two sentences. Here, the entire text represents a span for this technique, while the image is for reinforcement.

\textbf{~\\18. Bandwagon:}
Attempting to persuade the target audience to join in and take the course of action because ``everyone else is taking the same action.''

\begin{figure}[!tbh]
    \centering
    \includegraphics[width=\columnwidth]{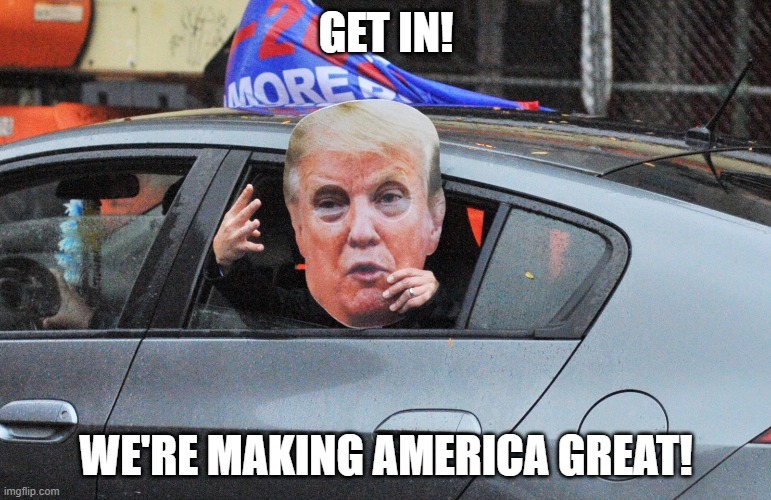}
    \caption{Example for \textbf{Bandwagon};\msrc
    \imgsrc{https://commons.wikimedia.org/wiki/File:3_Days_Until_Election_Day_-_Trump_Parade_and_Proud_Boys_(50558093408).jpg}{};
    \ccsnd{}
    }
    \label{fig:bandwagon_exmpl}
\end{figure}

Figure~\ref{fig:bandwagon_exmpl} shows an example that covers the entire text; the image less relevant.

\newpage
\textbf{~\\19. Smears:}
A smear is an effort to damage or to call into question someone's reputation, by propounding negative propaganda. It can be applied to individuals or groups. 

\begin{figure}[!tbh]
    \centering
    \includegraphics[width=\columnwidth]{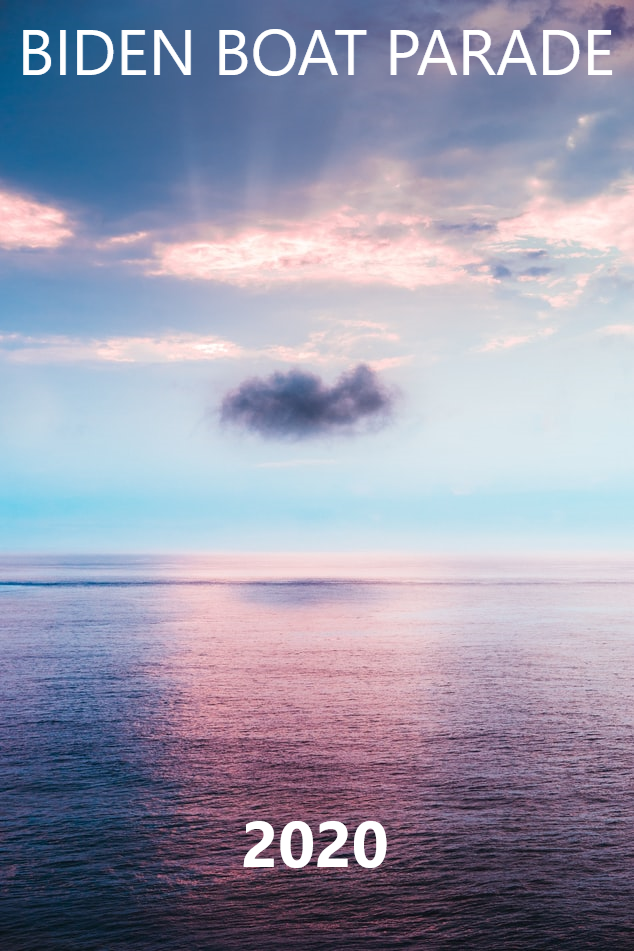}
    \caption{Example for \textbf{Smears};\msrc \imgsrc{https://unsplash.com/photos/xe-ss5Tg2mo}{};
    \unplash{}
    }
    \label{fig:smears_exmpl}
\end{figure}

An example meme is shown in Figure \ref{fig:smears_exmpl}, where the combination of the image and the text conveys the idea that Biden is unpopular.

\newpage
\textbf{~\\20. Glittering Generalities:}
These are words or symbols in the value system of the target audience that produce a positive image when attached to a person or issue. Peace, hope, happiness, security, wise leadership, freedom, ``The Truth'', etc. are virtue words. Virtue can be also expressed in images, where a person or an object is depicted positively.

\begin{figure}[!tbh]
    \centering
    \includegraphics[width=\columnwidth]{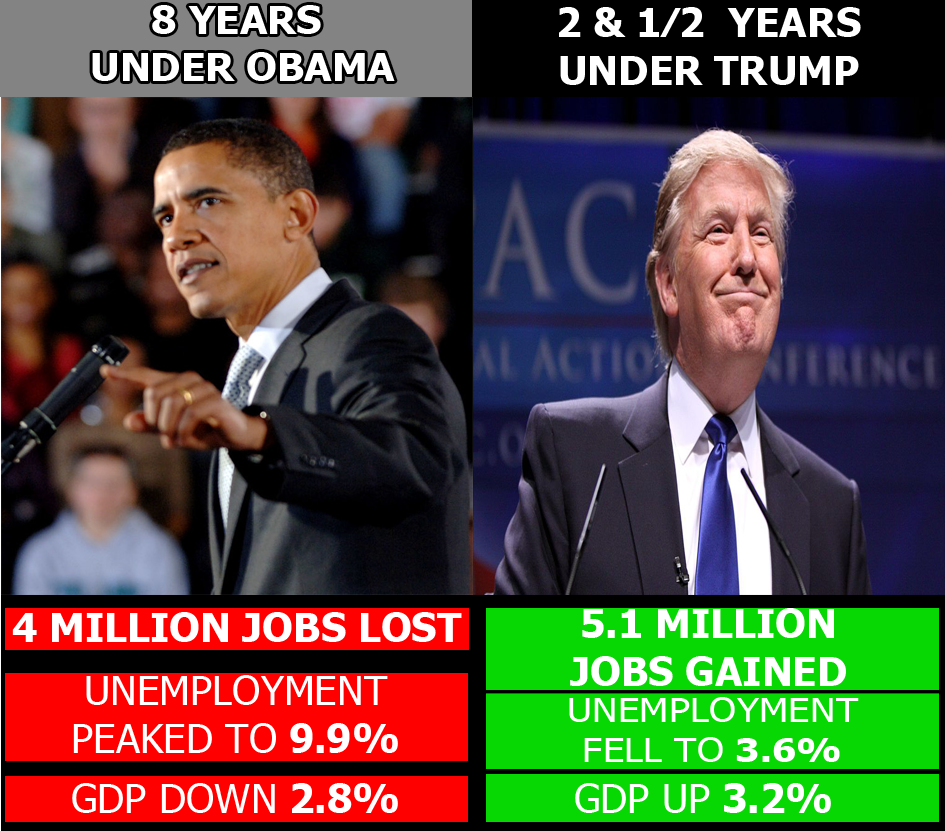}
    \caption{Example for \textbf{Glittering Generalities};\msrc \imgsrc{https://commons.wikimedia.org/wiki/File:Barack_Obama_(3619168415)_(cropped).jpg}{1},  \imgsrc{https://commons.wikimedia.org/wiki/File:Donald_Trump_(31963023360).jpg}{2}; 
    \ccsnd{1},
    \ccsnd{2}}
    \label{fig:glittering_exmpl}
\end{figure}

Figure \ref{fig:glittering_exmpl} shows an example of the use of this technique, in the right half of the meme. The technique covers the entire text span starting from ``\emph{2 \& 1/2 years} $\ldots$'' until ``\emph{GDP up 3.2\% $\ldots$}'' It is also expressed in the image, which depicts Donald Trump in a positive way. The text--image combination further strengthens the technique.

\newpage
\textbf{~\\21. Appeal to (Strong) Emotions:} 
Using images with strong positive/negative emotional implications to influence an audience. We reserve this technique to the images content only.

\begin{figure}[!tbh]
    \centering
    \includegraphics[width=0.98\columnwidth]{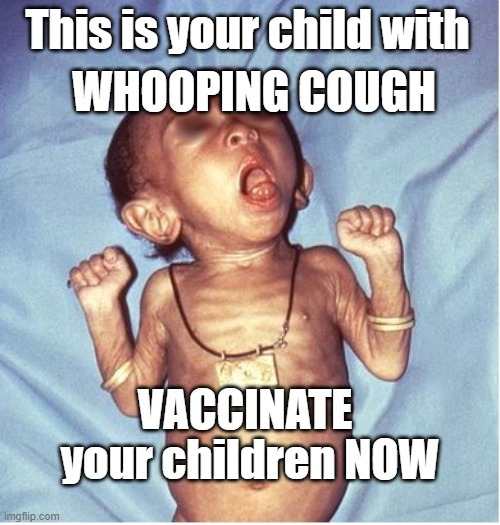}
    \caption{Example for \textbf{Appeal to (Strong) Emotions};\msrc \imgsrc{https://commons.wikimedia.org/wiki/File:Pertussis_lores.jpg}{};
    \public{}
    }
    \label{fig:strong_exmpl}
\end{figure}

An example is shown in Figure~\ref{fig:strong_exmpl}, which invokes strong emotions in the audience.

\newpage
\textbf{~\\22. Transfer:}
Also known as \emph{Association}, this is a technique of projecting positive or negative qualities (praise or blame) of a person, entity, object, or value onto another one to make the second one more acceptable or to discredit it. It evokes an emotional response, which stimulates the target to identify with recognized authorities. Often highly visual, this technique often utilizes symbols (for example, the swastikas used in Nazi Germany, originally a symbol for health and prosperity) superimposed over other visual images.

\begin{figure}[!tbh]
    \centering
    \includegraphics[width=\columnwidth]{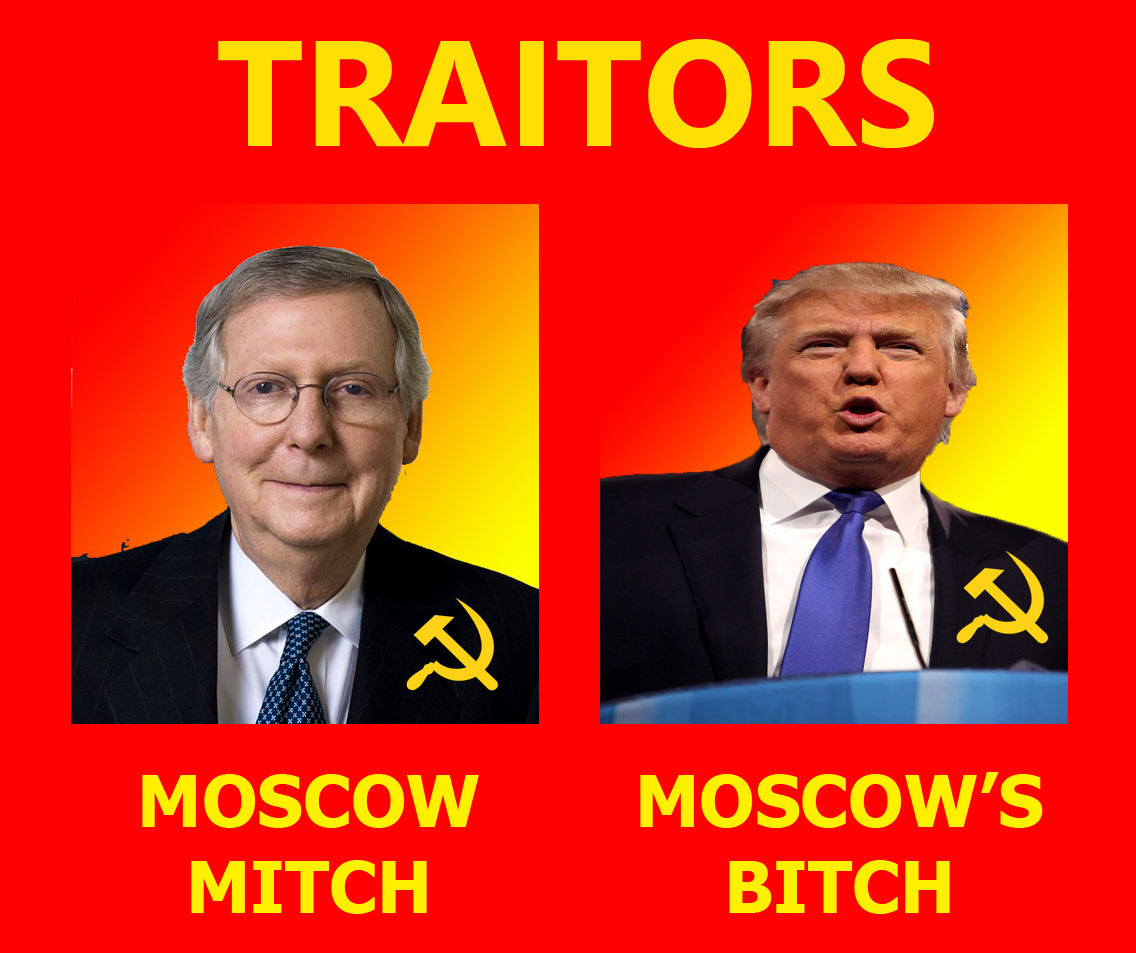}
    \caption{Example for \textbf{Transfer};\msrc
    \imgsrc{https://commons.wikimedia.org/wiki/File:Sen_Mitch_McConnell_official_(cropped).jpg}{1}, 
    \imgsrc{https://search.creativecommons.org/photos/ce37cc08-cf83-4a5b-9903-80d9e7d172e7}{2}; \public{1}, \ccsnd{2}
    }
    \label{fig:transfer_exmpl}
\end{figure}

Figure \ref{fig:transfer_exmpl} shows an example, where the \emph{Transfer} technique makes use of a communist symbol (namely, hammer and sickle) on top of the pictures of two targeted politicians, with the aim of depicting them in a negative way. The technique is further reinforced by the use of the red color (which is also a symbol of Communism), and by the two instances of \emph{Name Calling} (``\emph{Moscow Mitch}'' and ``\emph{Moscow's bitch}''), which make a connection to Moscow (which in turn was the capital of the former Communist block).

\clearpage
\newpage
\input{sections/tasks}

\clearpage
\newpage
\section{Participating Systems} 
\label{app:appendix_summary_sub_systems}

Below, we give a brief description of the participating systems, listed in alphabetical order, with reference to the corresponding task description paper. The numbers in square brackets refer to the official ranking of the target system on the individual subtasks.

\paragraph{1213Li~\cite{SemEval2021-6-Li}[ST3: 3rd]} used RoBERTa and ResNet-50 as feature extractors for texts and images. They used a label embedding layer with a multi-modal attention mechanism to measure the similarity between labels with the multi-modal information and fused features for label prediction.

\paragraph{AIMH~\cite{SemEval2021-6-AIMH} [ST1: 5th, ST3: 4th]} used transformer-based models and proposed visual--textual transformers to mainly address subtask 3 (ST3). For the visual part, they used ResNet50, and for the textual part, they used BERT. The same network used the multi-label classification on text (ST1) by using only the textual part of the network.

\paragraph{Alpha \cite{SemEval2021-6-alpha} [ST1:2nd, ST3:1st]} team pre-trained a transformer using text with visual features. They extract grid features, using ResNet50, and salient region features, using BUTD. They used grid features to capture the high-level semantic information found in the images. Additionally, they used salient region features to describe objects and to caption the event present in the memes. For ST1, they combined the text and the text representation of the visual features, and trained DeBERTa. For ST3, they built an ensemble of fine-tuned DeBERTA+ResNet, DeBERTA+BUTD, and ERNIE-VIL.

\paragraph{HOMADOS~\cite{SemEval2021-6-Kaczynski} [ST2: 2nd]} used a multi-task learning (MTL) approach with additional datasets such as the PTC corpus from SemEval-2020~\cite{da-san-martino-etal-2020-semeval}, and a fake news corpus \cite{Przybyla_2020}. The model was trained using BERT followed by several output layers, which solve auxiliary tasks of propaganda detection and credibility assessment in two distinct scenarios: sequential and parallel MTL, effectively accelerating the training process. The final submission used a parallel MTL approach on the propaganda detection of SemEval-2020, which ranked second.

\paragraph{TeamFPAI~\cite{SemEval2021-6-Hou} (ST2: 3rd)} formulated the task as a question answering one in a machine reading comprehension (MRC) framework, which achieved better results compared to an ensemble-based approach \cite{liu-etal-2018-multi}. Moreover, data augmentation and loss design techniques were also explored to alleviate the problem of data sparseness and imbalance. Their system was ranked 3rd in the ﬁnal evaluation phase.

\paragraph{CSECUDSG \cite{SemEval2021-6-Hossain} (ST1: 13th, ST2: 6th, ST3: 6th)} participated in all three subtasks. For ST1, they used a majority vote late fusion on top of logistic regression, decision tree, and fine-tuned DistilBERT models. For ST2, they reformulated the task as one of multi-label classification, where a pre-trained BERT model was used to design binary classifiers for each technique in a multi-label classification setting. For ST3, they used a majority voting late fusion on top of fine-tuned DistilBERT, ResNet50, and a predicted label from an early fusion model. The early fusion model consisted of features from (\emph{i})~multi-kernel CNN on top of the LSTM model with word embeddings including (\emph{ii})~word2vec \cite{mikolov2013efficient}, (\emph{iii})~word embeddings fine-tuned FastBERT \cite{liu-etal-2020-fastbert}, (\emph{iv})~RoBERTa, (\emph{v})~sentence embeddings from FastBERT, (\emph{vi})~image features from YouTube-8M \cite{abu2016youtube}, and (\emph{vii})~multimodal features from VisualBERT \cite{li2019visualbert}.

\paragraph{LeCun \cite{SemEval2021-6-LeCun} [ST1: 6th]} trained five models and combined them in an ensemble. Initially, they pre-processed text using stemming. Later, they trained DebERTA and RoBERTa models with augmented data using synonym replacement, random insertion, random swap, random deletion and back-translation. They first trained the five models separately, and then they fine-tuned the ensemble on the official non-augmented data.

\paragraph{LIIR~\cite{SemEval2021-6-Ghadery}[ST3: 8th]} used data augmentation through back-translation and CLIP to obtain image and text representations, which were then fed to a chained classifier that uses the correlations between the output techniques.  

\paragraph{LT3-UGent \cite{SemEval2021-6-Singh} [ST3: 14th]} participated in subtask 3 only. They used Multimodal Compact Bilinear Pooling to combine representations from ResNet-51 and BERT. They further fine-tuned on the PTC corpus~\cite{da-san-martino-etal-2020-semeval}.

\paragraph{MinD \cite{SemEval2021-6-Tian} [ST1: 1st, ST3: 2nd]} used five pre-trained models for \mbox{ST1}: BERT, RoBERTa, XLNet, DeBERTa, and ALBERT. They first fine-tuned them on the PTC corpus~\cite{da-san-martino-etal-2020-semeval}, and then on the training data. For the final prediction, they averaged the probabilities of the models. They also used a post-processing rule: a bigram that appeared more than three times was flagged as a \textit{Repetition}.
The system for \mbox{ST1} was also used for \mbox{ST3}, combined with (\emph{i})~ResNet-34, a face recognition system, (\emph{ii})~OCR-based positional embeddings for text boxes in the image, and (\emph{iii})~Faster R-CNN to extract region-based image features. They combined the textual and the visual representations by averaging their probabilities. Other multimodal fusion strategies included concatenation of the representation and mapping them to the space using a multilayer perceptron. 

\paragraph{NLP-IITR~\cite{SemEval2021-6-VGupta} [ST1: 15th]} used an ensemble that included included fine-tuned RoBERTa, BERT, and three additional models. They further used pre-processing. To tackle data scarceness for some rare labels, they used data augmentation using back-translation.

\paragraph{NLyticsFKIE~\cite{SemEval2021-6-NLyticsFKIE} [ST1: 9th, ST3: 13th]} used RoBERTa as a text encoder in ST1 and ST3. For ST1, they used RoBERTa's output to build a classifier to predict each label separately. For ST3, they still used RoBERTa to encode the text and a VGG-16 layer to encode the image. They used multiple copies of a cross-modality encoder that outputs an encoding of the image features with respect to the text features, and vice versa. The concatenation of the two cross-encoders' outputs was then passed through a residual layer followed by layer normalization.

\paragraph{Volta~\cite{SemEval2021-6-Volta} [ST1: 3rd, ST2: 1st, ST3: 5th]} used a combination of transformers for all subtasks. For ST1, they used RoBERTa's [CLS] token, which they fed to a feed-forward neural network, and example weighting to take care of class imbalance. For ST2, they predicted token classes by considering the output of each token embedding as obtained by RoBERTa. To account for subwords’ class, they merged each subword belonging to the same token and assigned the union of the subwords' labels.
For ST3, they separately encoded the textual features (extracted using RoBERTa) and the multi-modal features (extracted using UNITER, VisualBERT, and LXMERT). This layer's input was a sequence of textual subwords and visual tokens extracted by keeping the top 36 regions of interest as returned by Faster R-CNN. A concatenation of the two different [CLS] tokens was then fed into an MLP, and weighted labels were used with a cross-entropy loss.

\paragraph{WVOQ~\cite{SemEval2021-6-Roele} [ST2: 5th]} used a novel approach to ST2 consisting of adopting an encoder--decoder strategy. The encoder encodes the passage, while the decoder generates a marked version of the input, where the markup outlines the various spans along with the classes they belong to. In this way, the system performed simultaneous span detection and classification. The encoder--decoder used a specialization of BART.

\paragraph{YNU-HPCC~\cite{SemEval2021-6-YNUHPCC} [ST1: 12th, ST2: 7th, ST3: 11th]}
For ST1, they used a CNN on top of ALBERT and fine-tuned the model for multi-label classification. For ST2, each propaganda technique was considered as an independent task, and features were extracted from the pre-trained BERT model. Subsequently, the problem was addressed as a multi-task sequence labeling one, and the results for each task were combined. For ST3, a multi-modal network was used, where embeddings from textual and visual networks were concatenated, which was followed by a fully connected layer. For the text, the same approach was used for ST1, and for the image, ResNet and VGGNet were used for image feature extraction.

%% file: sections/tasks.tex
\section{Subtasks: Definition, Data Format, and Data Examples}
\label{sec:tasks}

Below, we describe the three subtasks and the general data format for each of them. We further show an example of an annotated example for each subtask.

\subsection{Subtask 1}
This is a multi-label classification problem, defined as follows:

\begin{description}
\item \textbf{Subtask 1 (ST1)} Given only the ``textual content'' of a meme, identify which of the 20 techniques are used in it. 
\end{description}

The data for ST1 comes as a JSON object in the following format:

{\small
\begin{verbatim}
{
  id -> example identifier,
  labels -> list of persuasion
            techniques,
  text -> text of the meme
}
\end{verbatim}
}

Here is an example:
{\small
\begin{verbatim}
{
 "id": "125",
 "labels": [
    "Loaded Language",
    "Name calling/Labeling"
 ],
 "text": "I HATE TRUMP\n\n
          MOST TERRORIST DO"
}
\end{verbatim}
}

\newpage
\subsection{Subtask 2}

ST2 is a more complex version of ST1, as it asks not only for the techniques but also for the exact spans of use each technique.
This subtask is a combination of the two subtasks in \textit{SemEval-2020 task 11}. It is a multi-label sequence tagging problem, defined as follows:
 
\begin{description}
\item \textbf{Subtask 2 (ST2)} Given only the ``textual content'' of a meme, identify which of the 20 techniques are used in it together with the span(s) of text covered by each technique.
\end{description}

The data for ST2 comes as a JSON object with the following format:

{\small
\begin{verbatim}
{
  id -> example identifier,
  text -> text of the meme
  labels : [ -> list of objects
    {
      start -> start index,
      end -> end index,
      technique -> technique,
      text_fragment -> text
    }
  ]
}
\end{verbatim}
}

Here is an example:

{\small
\begin{verbatim}
{
 "id": "125",
 "text": "I HATE TRUMP\n\n
          MOST TERRORIST DO"
 "labels": [
   {
    "start": 2,
    "end": 6,
    "technique": "Loaded Language",
    "text_fragment": "HATE"
   },
   {
    "start": 19,
    "end": 28,
    "technique": "Name calling/
    Labeling",
    "text_fragment": "TERRORIST"
   }
 ]          
}
\end{verbatim}
}
Note that the labels to be predicted for ST2 are the same ones as for ST1, but this time the spans are to be predicted as well.

\subsection{Subtask 3}

ST3 is a multi-modal version of ST1, where the image is also provided. It is a multi-label classification problem, defined as follows:

\begin{description}
\item \textbf{Subtask 3 (ST3)} Given a meme, identify which of the 22 techniques are used both in the textual and in the visual content of the meme.
\end{description}

The data for ST3 comes as a JSON object with the following format:

{\small
\begin{verbatim}
{
  id -> example identifier,
  labels -> list of persuasion
            techniques,
  image -> name of the image file,
  text -> text of the meme
}
\end{verbatim}
}
Here is an example:

{\small
\begin{verbatim}
{
 "id": "125",
 "labels": [
    "Loaded Language",
    "Name calling/Labeling",
    "Reductio ad hitlerum",
    "Smears",
 ],
 "image": "125_image.png"
}
\end{verbatim}
}

\begin{figure}[tbh]
    \centering
    \includegraphics[width=\columnwidth]{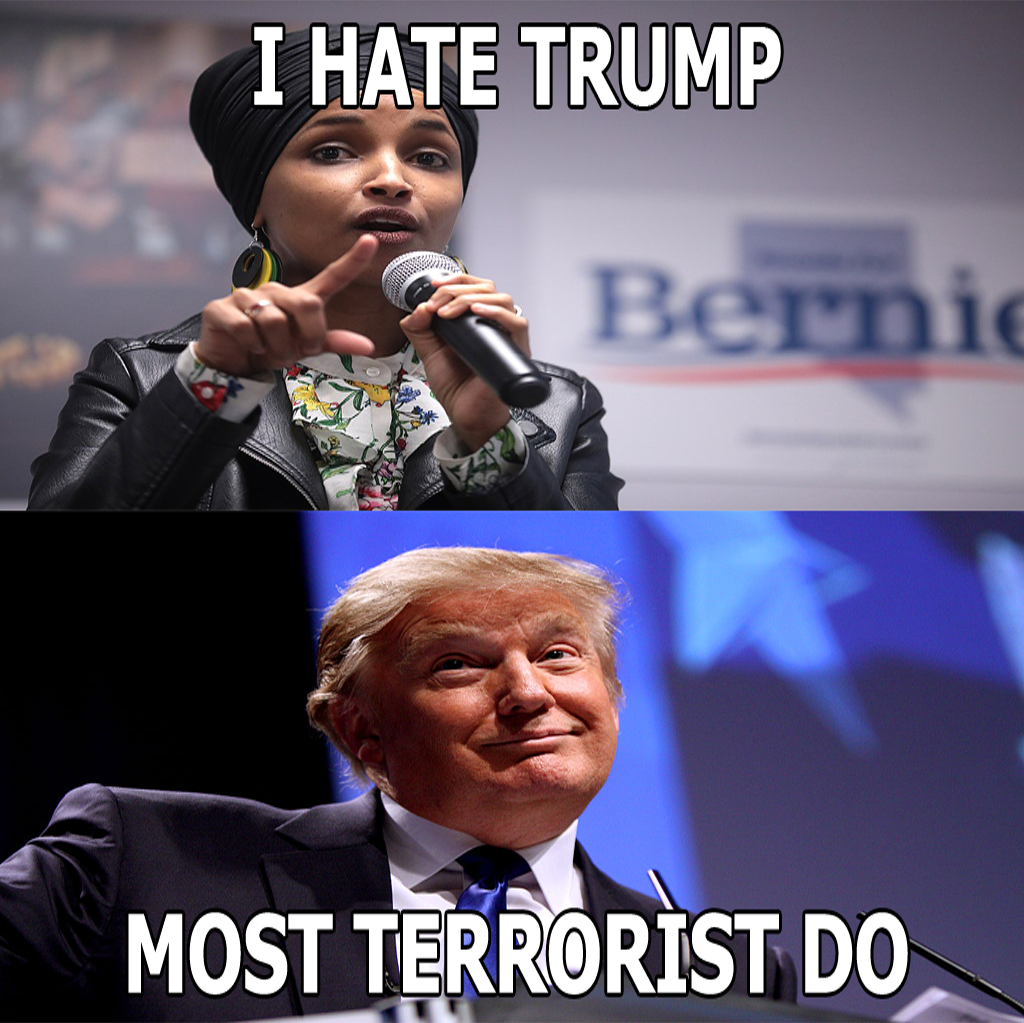}
    \caption{The meme with id=125; \msrc
    \imgsrc{https://commons.wikimedia.org/wiki/File:Ilhan_Omar_(49518038586).jpg}{1},
    \imgsrc{https://commons.wikimedia.org/wiki/File:Donald_Trump_by_Gage_Skidmore_2.jpg}{2};
    \ccsnd{1}, \cctrdunprt{2}}
    \label{fig:hitlerum}
\end{figure}

Here, the image, which is shown in Figure~\ref{fig:hitlerum}), gives rise to two additional persuasion techniques compared to ST1: \emph{Reductio ad Hitlerum} and \emph{Smears}. These techniques are not clearly present in the text alone. Indeed, the image is needed for us to see that there is \emph{Smears}, as this can be only seen when we understand that this is a dialog with a negative propaganda targeting one of the participants (Ilhan Omar). Similarly, we need the image for \emph{Reductio ad Hitlerum}: the image shows us that Ilhan Omar is depicted as a bad person (she is targeted by the \emph{Name Calling} ``\emph{terrorist}'', and she is also the target of the \emph{Smears}), and thus the message being conveyed is that any choice that such a bad person does has to be a bad choice, i.e., hating Trump is a bad thing to do as this is something terrorists do.

